\documentclass[%
 reprint,
 amsmath,amssymb,
 aps,
floatfix,
]{revtex4-2}
\usepackage{ulem}
\usepackage{graphicx}
\usepackage{graphicx}
\usepackage{bm}
\usepackage{hyperref}
\usepackage{amsmath}
\usepackage{braket}
\usepackage{siunitx}
\usepackage{mathtools}
\usepackage{amssymb} 
\usepackage{mathrsfs} 
\usepackage{lipsum}
\usepackage{float}
\usepackage{booktabs}
\usepackage[section]{placeins}
\usepackage{ragged2e} 
\usepackage{caption}

\DeclareCaptionJustification{justified}{\justifying}
\begin{document}

\preprint{}

\title{Modeling Quantum Optomechanical STIRAP}

\author{Ian Hedgepeth$^1$}
\author{Youqiu Zhan$^1$}
\author{Vitaly Fedoseev$^2$}
\author{Dirk Bouwmeester$^{1,2}$}

\affiliation{$^1$Department of Physics, University of California, Santa Barbara, California 93106, USA}
\affiliation{$^2$Huygens-Kamerlingh Onnes Laboratory, Universiteit Leiden, 2333 CA Leiden, The Netherlands}

\date{\today}

\begin{abstract}
 
Quantum optomechanical STIRAP (Stimulated Raman Adiabatic Passage) is investigated for a system of two mechanical modes coupled to an optical mode. We show analytically that in a system without loss, fractional STIRAP can generate a mechanical Bell state from a single phonon Fock state of one of the mechanical modes with the other mechanical mode in the vacuum state, and a product state from a coherent state. Relative phases between Fock basis components in the final state of STIRAP are determined by the phonon-number parity of the initial state. Furthermore, the system is numerically studied to determine the effects of dissipation, and it is concluded that high-fidelity entanglement can be achieved via fractional STIRAP using state-of-the-art cryogenic cooling and mechanical devices. Finally, an interferometric protocol using time-reversed fractional STIRAP is proposed to quantify entanglement between two mechanical modes.
 
\end{abstract}
\maketitle
\section{Introduction}
Quantum state transfer and generating quantum entanglement are essential to the field of quantum information science. In the case of quantum states of mechanical devices, such as membranes \cite{OG_membranes}, phononic crystals \cite{phononic_materials} and phononic cavity structures \cite{ARJUNAN20221}, the coupling between modes is often achieved via intermediate optical modes \cite{interferometry}. The fidelity of state transfer between mechanical modes with this method is limited by the mechanical quality factors and loss from the optical mode \cite{Optically_Mediated_dissipation}. 

Typically, the optical mode is engineered with a low-quality factor compared to the mechanical modes ($Q_\mathrm {opt} \ll Q_\mathrm m$) to facilitate efficient external coupling; however, this creates a potential decoherence channel 
\cite{Wang_2017}\cite{Laser_Cooling}.
To circumvent this, Stimulated Raman Adiabatic Passage (STIRAP), initially developed in the context of atomic physics, has been adapted to the optomechanical domain \cite{Wang_2017} \cite{Ian's_Paper}. STIRAP is a coherent process that enables adiabatic state transfer while ensuring the intermediary state—the optical cavity mode—remains unpopulated \cite{stirap}. By maintaining the system in such a “dark state,” the transfer becomes robust against optical decay, with the final fidelity limited only by the dissipation and decoherence of the initial and target mechanical modes. 

While the effects of dissipation on STIRAP have been extensively modeled in cavity QED \cite{Wang_2017}, recent work has demonstrated its efficacy in the classical regime of optomechanics \cite{Ian's_Paper} and through numerical modeling of multi-mode systems \cite{multimode}. Furthermore, by precisely tailoring the driving fields, one can implement fractional STIRAP (fSTIRAP) \cite{fSTIRAP} to prepare a coherent superposition of mechanical states. 

In this article, optomechanical STIRAP in the quantum regime is investigated with the aim to create and characterize entangled states between mechanical oscillators.
To study the dynamics of a multi-mode optomechanical system undergoing STIRAP, section 2 presents an analytic model for such a system. An important finding is that phonon Fock states acquire relative phases upon (f)STIRAP, determined by the phonon number parity. Section 3 presents a numerical study using a Lindblad master equation, which is solved using the Quantum Toolbox in Python (QuTiP) \cite{QuTiP}. Optomechanical (f)STIRAP under various scenarios and realistic environmental conditions is considered. Additionally, it is shown that through the use of fSTIRAP and different initial states, a mechanical Bell state and a product of coherent states can be generated. 
Finally, it is shown that a test of entanglement can be performed through applying a reverse fSTIRAP pulse sequence. 

\section{Analytical study of optomechanical STIRAP}
\label{analytical}

STIRAP has initially been proposed in the context of three-level $\Lambda$-type systems \cite{stirap_shore}.
The Hilbert space of a generic optomechanical system consists however of an optical cavity mode (c) and two mechanical modes, and is the tensor product of three Fock spaces. An orthonormal basis exists labeled by the number of bosons,
denoted as $\left|n_\mathrm c,n_1,n_2\right>$.
To apply the concept of STIRAP, consider a multi-mode optomechanical Hamiltonian \cite{multimode} in a frame rotating with respect to the optical cavity and mechanical modes,
\begin{equation}
\label{triply}
\begin{split}
\hat H = \sum_{i,j=1}^2 G_{i,j}(t) &\left(\hat{a}^\dagger e^{i{\Delta_i}t} + \hat{a} e^{-i{\Delta_i}t}\right) \\
\times&\left(\hat{b}_j e^{-i\omega_j t} + \hat{b}_j^\dagger e^{i\omega_j t}\right),
\end{split}
\end{equation}
where $G_{i,j}(t) = g_j\alpha_i(t)$.
Here $g_j$ are the single-photon optomechanical coupling strengths associated with each mechanical mode;
$\alpha_i(t)$ are the time-dependent optical driving field strengths \cite{multimode}, $\hat{a}^\dagger$($\hat{a}$) is the creation (annihilation) operator associated with the optical cavity;
$\Delta_i=\omega_\mathrm c - \omega_{\mathrm Li}$ is the detuning of the pump (with frequency $\omega_{\mathrm Li}$) with respect to the cavity frequency ($\omega_\mathrm c$);
$\hat{b}^\dagger_i$ ($\hat{b_i}$) are the creation (annihilation) operators associated with the mechanical mode $i$;
and $\omega_j$ are the mechanical mode frequencies.

Making the experimentally realistic assumptions that $|\omega_1\pm\omega_2|$
is much larger than both $|\Delta_i-\omega_i|$ ($i=1,2$)
and $1/\sigma_G$, where $\sigma_G$ is the time scale over which $G_{i,j}$ varies,
the (triple) rotating-wave approximation can be applied
to drop fast-oscillating terms in the Hamiltonian. This yields: 
\begin{equation}
\label{hamiltonian rwa}
    \hat H=\sum_{i=1}^2 G_{i,i}(t)\left(\hat a^\dagger \hat b_i e^{i\phi_i(t)}+\hat a \hat b_i^\dagger e^{-i\phi_i(t)}\right),
\end{equation}
where $\phi_i(t)=\left(\Delta_i-\omega_i\right)t$.

After defining
\begin{equation}
\label{b-}
\hat b_-(t)=\frac{G_{2,2}(t)\,\hat b_1e^{i\phi_1(t)}-G_{1,1}(t)\,\hat b_2e^{i\phi_2(t)}}{\sqrt{G_{1,1}(t)^2+G_{2,2}(t)^2}},
\end{equation}
one can verify that $\left[\hat b_-,\hat b_-^\dagger\right]=1$ in order to see that it is a boson annihilation operator.
Physically, it represents a collective mode that involves both mechanical modes.
Using this, one can define a state for each $n$,
\begin{equation}
\left|\Phi_n(t)\right>=\frac1{\sqrt{n!}}\hat b_{-}^{\dagger}(t)^{n}\left|0\right>,
\end{equation}
which is a linear combination of states $\left|0,n_1,n_2\right>$ satisfying $n_1+n_2=n$.
Noticing that $\left[\hat H,\hat b_-\right]=0$, we have
$\hat H\left|\Phi_n\right>=(\hat b_-^{\dagger})^n\hat H\left|0\right>/\sqrt{n!}=0$,
so $\left|\Phi_n\right>$ is an eigenstate of $H$ with zero energy.

If the system is initially in this state,
assuming that  $G_{1,1}(t)$ and $G_{2,2}(t)$ change slowly with time, the system will stay in this state.
This is due to the adiabatic theorem for the case of $n=1$,
and Appendix~\ref{energy gap dark state} provides a justification of this claim for more general cases.
Because $\hat a\left|\Phi_n\right>=0$, the system does not populate the cavity mode throughout the adiabatic passage,
which means that $\left|\Phi_n\right>$ is the dark state.

The mixing angle $\vartheta (t)$ can be defined by $\tan\vartheta(t)=G_{1,1}(t)/G_{2,2}(t)$,
and examined to see how $\hat{b}_-(t)$ changes when $\vartheta(t)$ varies adiabatically.
Because, $\hat b_-(\vartheta=0)=\hat b_1e^{i\phi_1}$ and $\hat b_-(\vartheta=\pi/2)=-\hat b_2e^{i\phi_2}$,
$\left|\Phi_n(\vartheta=0)\right>=e^{in\phi_1}\left|0,n,0\right>$
and $\left|\Phi_n(\vartheta=\pi/2)\right>=(-1)^ne^{in\phi_2}\left|0,0,n\right>$.
Therefore, if the system is initially in the state $\left|0,n,0\right>$,
when $\vartheta$ is varied adiabatically from $0$ to $\pi/2$,
the system will evolve to the state $\left|0,0,n\right>$ up to an overall phase, describing optomechanical STIRAP.
If $\vartheta$ changes adiabatically from $0$ to some general angle $\theta$,
the system will evolve to some superposition state of $n+1$ Fock states, where the weighting will depend on $\theta$. This general case can be referred to as fractional STIRAP.

Formally, the transfer of $n$ phonons in the first mechanical mode to $n$ phonons in the second mechanical mode,
elaborated above,
is a STIRAP-like process on a multistate chain of $2n+1$ states.
The standard Hamiltonian for such a process is \cite{stirap}
\begin{equation}
\hat H = \frac{1}{2} \begin{pmatrix}
0 & \Omega_1 e^{-i\phi_1} & & \\
\Omega_1 e^{i\phi_1} & 0 & \ddots & \\
& \ddots & \ddots & \Omega_{2n} e^{-i\phi_{2n}} \\
& & \Omega_{2n} e^{i\phi_{2n}} & 0
\end{pmatrix}
\label{ham}
\end{equation}
(explicit time dependence is suppressed to save space).
Equation \eqref{ham} is obtained from ~\eqref{hamiltonian rwa} by truncating its matrix form
to only include the rows and columns corresponding to the states $\left|n_\mathrm c,n_1,n_2\right>$
that satisfy $n_\mathrm c+n_1+n_2=n$ and $n_\mathrm c\le 1$.
Comparing it with the standard form gives
\begin{equation}
\begin{gathered}
  \Omega_{2k-1}=2\sqrt{n-k+1}\,G_{1,1}, \quad \phi_{2k-1}=\phi_1, \\
  \Omega_{2k}=2\sqrt{k}\,G_{2,2}, \quad \phi_{2k}=\phi_2. \\
  (k=1,\dots,n)
\end{gathered}
\end{equation}
The dynamics of the optomechanical system is exactly the same as that of the standard multistate STIRAP when $n\le1$.
For other $n$, they are not exactly the same because the subspace spanned by those states is not an invariant subspace of $\hat H$.
However, they are the same in the adiabatic limit because the dark state has populations only in the basis states with $n_\mathrm c=0$,
and the matrix elements of $\hat H$ with those states and states truncated away are zero. From now on, except when stated otherwise, resonant driving of the two transitions is considered, and thus, it is assumed that $\Delta_i=\omega_i$.

An interesting special case to consider is $n=1$, where the transfer reduces to the standard 3-state (fractional) STIRAP~\cite{stirap_shore}.
In this case, the dark state is
$\ket{\Phi_1}=\cos\vartheta\ket{0,1,0}-\sin\vartheta\ket{0,0,1}$.
Specifically, after the fractional STIRAP with the final mixing angle $\theta=\pi/4$,
the system will end up in the Bell state $\ket{\Psi^-}=\left(\ket{0,1,0}-\ket{0,0,1}\right)/\sqrt2$.

Additionally, the transfer of a superposition of different numbers of phonons can be studied.
Neither the dynamic phase nor the geometric phase is generated as $\left|\Phi_n\right>$ changes over time.
The dynamic phase is zero because the energy of this state is fixed to zero, and
the geometric phase is zero because this state is always a real linear combination of time-independent basis vectors.
Therefore, when the system starts with a linear combination of phonon Fock states
in the first mechanical mode,
each component evolves without developing relative phases besides the phases resulting from \eqref{b-}.
In other words, after fractional STIRAP with an arbitrary final mixing angle $\theta$,
the initial state $\sum_nc_n\left|0,n,0\right>$ evolves to $\sum_nc_n\left|\Phi_n(\vartheta=\theta)\right>$.
Specifically, for STIRAP, the initial and final states can be written as
\begin{equation}
\centering
   \sum_nc_n\left|0,n,0\right> \rightarrow \sum_n(-1)^nc_n\left|0,0,n\right>,  \quad \theta=\pi / 2
\end{equation}
showing that the state evolution for optomechanical STIRAP is dependent on the parity of $n$.
This can be further specified for the case of a single phonon, $\left(\ket{0,0,0}+\ket{0,1,0}\right)/\sqrt2$ evolves into $\left(\ket{0,0,0}-\ket{0,0,1}\right)/\sqrt2$ after STIRAP.

Another interesting initial state to explore is the coherent state, $\ket{\alpha}_1=e^{\alpha\hat b_1^\dagger-\alpha^*\hat b_1}\left|0\right>$. 
It evolves through adiabatic passage to $e^{\alpha\hat b_-(t)^\dagger-\alpha^*\hat b_-(t)}\left|0\right>$.
Because $\hat b_1$ and $\hat b_2$ commute, after fSTIRAP with mixing angle $\theta$,
the system becomes $e^{\alpha\hat b_1^\dagger\cos\theta-\text{h.c.}}e^{-\alpha\hat b_2^\dagger\sin\theta-\text{h.c.}}\left|0\right>=\ket{\alpha\cos\theta}_1\ket{-\alpha\sin\theta}_2$,
which is the tensor product of a coherent state in the first mechanical mode
and a coherent state in the second mechanical mode. This implies that
the two mechanical modes cannot be entangled in the final state, regardless of the value of $\theta$.

To quantify entanglement in the final state, the Negativity $\mathcal{N}(\rho_{12})$ is considered \cite{Negativity_entangled}. This measure is based upon the Peres--Horodecki criterion \cite{Peres_Horo}, which established that a bipartite system is entangled if its partial transpose possesses at least one negative eigenvalue. In the case of an optical cavity and two mechanical modes, the optical mode must first be traced over to reduce the system to bipartite. The resulting reduced density matrix, $\rho_{12}$ can then be compared to its partial transpose $\rho^{\mathrm T_2}_{12}$. This can be done through defining $\mathcal{N}(\rho_{12})$ in terms of $\rho^{\mathrm T_2}_{12}$ as,
\begin{equation}
    \label{Negativity}
    \mathcal{N}(\rho_{12})=\frac{\left\|\rho_{12}^{\mathrm T_2}\right\|_1-1}{2}
\end{equation}
where $\left\|\rho_{12}^{\mathrm T_2}\right\|_1 = \text{Tr}\sqrt{(\rho_{12}^{\mathrm T_2})^\dagger \rho_{12}^{\mathrm T_2}}$ denotes the trace norm, which is equivalent to the sum of the singular values of the partial transpose. In the case of fSTIRAP, where the result of $n=1$ is $\ket{\Psi^-}=\left(\ket{0,1,0}-\ket{0,0,1}\right)/\sqrt2$, $\mathcal{N}(\rho_{12})=1/2$, determined by $\rho^{\mathrm T_2}_{12}$ having eigenvalues of $\lambda=\pm1/2$. 

We now consider the specific form that the time-dependent functions $\alpha_i(t)$ can take to implement the adiabatic change in $\vartheta$.
For STIRAP, $\vartheta(t=-\infty)=0$
and $\vartheta(t=\infty)=\pi/2$, which is achieved by letting $\alpha_i$ take the form of two Gaussians,
\begin{equation}
\label{stirap pumps}
\alpha_1(t) = \frac{\Omega_0}{2g_1} e^{-(t - \tau)^2/\sigma^2}, \quad \alpha_2(t) = \frac{\Omega_0}{2g_2} e^{-(t + \tau)^2/\sigma^2},
\end{equation}
where $\tau$ is the separation between pulses, and $\sigma$ is the pulse width. 
The pulse amplitude $\Omega_0/2g_i$ represents the peak intracavity coherent field amplitude, and is related to the maximum mean photon number $\bar{n}_i=(\Omega_0/2g_i)^2$ in the $i$th cavity mode.

For fractional STIRAP with a final mixing angle $\theta$, the following can be used \cite{fSTIRAP}
\begin{equation}
\label{fractional pumps}
\begin{split}
\alpha_1(t) &= \frac{\Omega_0}{2g_1} \sin(\theta) \,e^{{-(t - \tau)}^2/\sigma^2},\\
\alpha_2(t) &= \frac{\Omega_0}{2g_2} e^{{-(t + \tau)}^2/\sigma^2} + \frac{\Omega_0}{2g_2} \cos(\theta) \,e^{-{(t - \tau)}^2/\sigma^2}.
\end{split}
\end{equation}
The adiabaticity condition expressed in terms of the parameters $\theta,\tau,\sigma$ is derived in Appendix~\ref{section adiabaticity condition}.
From now on, except when stated otherwise,
it is assumed that $g_1=g_2$ and $\alpha_0=\Omega_0/2g_1=\Omega_0/2g_2$.

\section{Simulation of STIRAP with decoherence and dissipation}
\subsection{STIRAP}
\label{STIRAP}

In this section, we simulate STIRAP in optomechanics, taking into account phonon/photon loss and coupling to a thermal bath. A generic method to include decoherence and dissipation in an open quantum system, consisting of an optical mode and two mechanical modes coupled to a thermal bath, is through the Lindblad Master Equation with the Hamiltonian found in \eqref{triply},

\begin{multline}
\label{master}
\dot{\hat \rho}(t) = -i\left[\hat H,\hat\rho\right] + \kappa \mathcal{L}(\hat{a})
\\+ \sum_{i=1}^2 \Gamma_{\mathrm i}\Big{(}(\bar{n}_i + 1) \mathcal{L}(\hat{b}_i) + \bar{n}_i\mathcal{L}(\hat{b}_i^\dagger)\Big{).}
\end{multline}
Here $\mathcal{L}$ is the Lindblad superoperator, which describes the non-unitary evolution of the system and can be written for a general operator $\hat{{C}}$ as $\mathcal{L}(\hat{C})\hat{\rho} = \hat{C}\hat{\rho}\hat{C}^\dagger - \frac{1}{2}\{\hat{C}^\dagger\hat{C}, \hat{\rho}\}$\cite{Lindbladian}. We consider photon leakage from the cavity at a rate $\kappa$, mechanical damping, represented by the linewidth of the mechanical resonator $\Gamma_i$ \cite{Steeneken_2021}, and thermalization via an external bath, modeled as the Bose--Einstein distribution \cite{Optomechanics} with $\bar{n}_{\mathrm i} =\left(e^{\omega_i/k_\mathrm BT}-1\right)^{-1}$ the average thermal population of the environment in terms of its temperature $T$, mechanical frequency $\omega_i$ and Boltzmann's constant $k_\mathrm B$.

To numerically simulate STIRAP, QuTiP is utilized to model \eqref{master} using the pulses from \eqref{stirap pumps} \cite{QuTiP}. We focus on the state transfer between two mechanical modes mediated by a cavity state, using the parameters listed in Table~\ref{tab:stirap_parameters}, assuming state-of-the-art resonators with $\text{Q}_i= \omega_i / \Gamma_i = 10^9$, and state-of-the-art cryogenic cooling \cite{Laser_Cooling,HOJ,Measurement_Motional}. These parameters are found by using the analysis of the parameter space in Appendix \ref{Dependence and Optimization of Parameters}, alongside setting realistic values using experimental parameters \cite{Ian's_Paper}. Using these parameters, the Hamiltonian in \eqref{triply}, and the master equation from \eqref{master}, three cases are considered.
\begin{table}[htbp]
\centering
\caption{Parameters for Numerical Simulation of STIRAP}
\label{tab:stirap_parameters}
\begin{tabular}{lll}
\toprule
\textbf{Parameter} & \textbf{Symbol} & \textbf{Value} \\
\midrule
Mechanical Frequency 1 & $\omega_1 / 2\pi$ & \SI{1.2}{MHz} \\
Mechanical Frequency 2 & $\omega_2 / 2\pi$ & \SI{1.8}{MHz} \\
Mechanical Quality Factor & $Q_1 , Q_2$ & \num{e9} \\
\midrule
Cavity Frequency & $\omega_\mathrm c / 2\pi$ & \SI{540}{THz} \\
Optical Decay Rate & $\kappa / 2\pi$ & \SI{2}{kHz} \\
Optomechanical Coupling & $g_{1} / 2\pi, g_{2} / 2\pi$ & \SI{2.5}{Hz} \\
\midrule
Pulse Width & $\sigma_1, \sigma_2$ & \SI{0.60}{ms} \\
STIRAP Pulse Delay & $\tau$ & $\sigma_1 / 1.43$ \\
fSTIRAP Pulse Delay & $\tau_f$ & $\sigma_1 / 1.25$ \\
Pump Amplitude & $\alpha_0$ & 2000 \\
\bottomrule
\end{tabular}
\end{table}

\begin{figure}[t]
    \centering
    \includegraphics[width=0.99\linewidth]{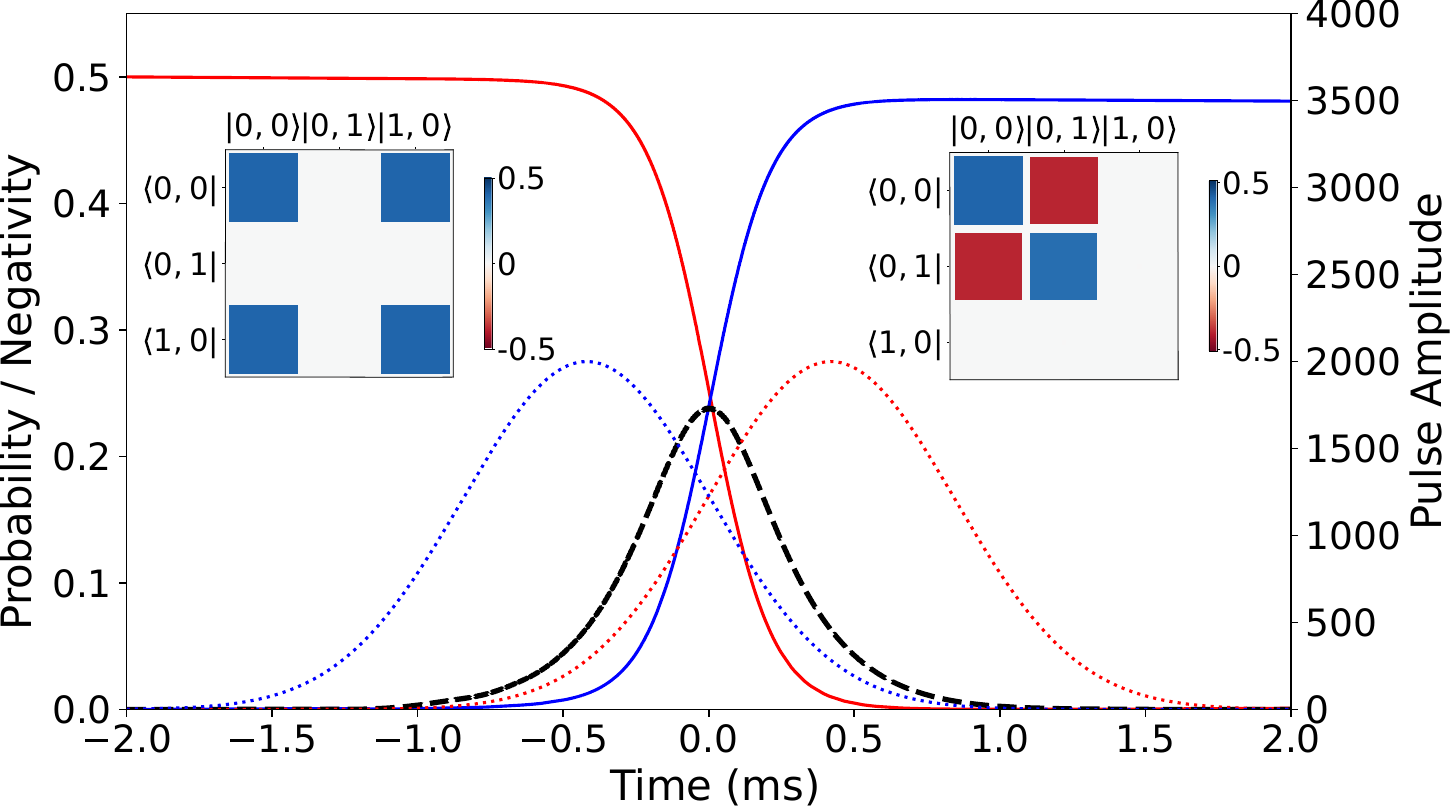}
    \caption{Time evolution of STIRAP of the superposition state $\frac{1}{\sqrt{2}}(\ket{0}_1+\ket{1}_1)$ at \SI{10}{mK}. The solid curves track the single phonon probability amplitudes for the first mechanical resonator (red) and second mechanical resonator (blue). The dotted Gaussian profiles represent the counterintuitive pulse sequence, where the Stokes pulse (red) leads the pump pulse (blue). Inset shows Hinton plots of the truncated reduced density matrices $\rho_{12}$ for the initial (left) and final (right) states. The black dashed line shows the entanglement evolution quantified by the negativity $\mathcal{N}(\rho_{12})$, which reaches a peak value of $\mathcal{N} \approx 0.25$ during maximal pulse overlap at \SI{0}{ms}. The state transfer at \SI{2.0}{ms} achieves a fidelity of 0.98 when measured against the state $\frac{1}{\sqrt{2}}(\ket{0}_2-\ket{1}_2)$.}
    \label{STIRAP with state 0+1 at 0.01k}
\end{figure}

In case 1, a bath temperature of \SI{10}{mK} is considered, and STIRAP is simulated using the initial state, $\frac{1}{\sqrt{2}}(\ket{0}_1+\ket{1}_1)\otimes \ket{0}_2 \ket{0}_c$, where mechanical mode 2 and the cavity are initially in the vacuum state, which is shown in Figure~\ref{STIRAP with state 0+1 at 0.01k}. Henceforth, the notation $\frac{1}{\sqrt{2}}(\ket{0}_i+\ket{1}_i)$, $i = \{1, 2\}$, can be used as shorthand for the state, with the other mechanical mode and cavity mode in the vacuum state. The pulses for STIRAP, from \eqref{stirap pumps} are shown (dotted), and it can be seen, by examining the reduced density matrix $\rho_{12}$ of mechanical modes 1 and 2, that the state has undergone a relative phase change and has become, $\sim\frac{1}{\sqrt{2}}(\ket{0}_2-\ket{1}_2)$ with fidelity 0.98, closely matching what was found in Section \ref{analytical}. The results in Figure \ref{STIRAP with state 0+1 at 0.01k} includes non-ideal adiabatic following and a change in single phonon population due to thermalization to the mean temperature of the bath ($\bar{n}_i\approx173$).

The degree of entanglement between the mechanical modes generated during the STIRAP process can be identified by observing the negativity $\mathcal{N}(\rho_{12})$ associated with the system \cite{Negativity_entangled}. The negativity of this reduced density matrix for \SI{10}{mK} case is shown in Figure \ref{STIRAP with state 0+1 at 0.01k} as the black dashed line. The negativity rises sharply during pulse overlap, peaking at a value of $\mathcal{N} \approx 0.25$. This peak corresponds to the point of maximal entanglement of the single-phonon excitation between the two modes. Notably, the maximum value is bounded at 0.25 (rather than 0.5 for the case of an initial state of a single phonon) because the initial state, $\frac{1}{\sqrt{2}}(\ket{0}_1+\ket{1}_1)$, contains a significant vacuum component that can be decoupled, as it does not share coherence with the single excitation entanglement, and excluded from the contribution to the bipartite entanglement.

In case 2, an initial state of a heralded single phonon state can be considered; see Appendix~\ref{experimental appendix},
\begin{equation}
\label{fstirap_real_initial}
\begin{aligned}
    \rho_{1}^\mathrm{i} &= 0.89\ket{1}_1\bra{1}_1 + 0.10\ket{2}_1\bra{2}_1 + 0.01\ket{3}_1\bra{3}_1. \\
\end{aligned}
\end{equation}
Using the state in \eqref{fstirap_real_initial}, STIRAP is simulated at \SI{50}{mK} and it is found that the state transfer can achieve a fidelity of 0.93, when compared to $\rho_2^\mathrm{f} = 0.89\ket{1}_2\bra{1}_2 + 0.10\ket{2}_2\bra{2}_2 + 0.01\ket{3}_2\bra{3}_2 $. At this temperature, the fidelity of state transfer decreases as the effects of thermalization have become stronger, due to a larger $\bar{n}_\mathrm b\approx867$.

In case 3, we consider STIRAP, for the case where the temperature has increased to \SI{1}{K} ($\bar{n}_\mathrm i\approx17000$) and is simulated using the state shown in \eqref{fstirap_real_initial}. It is found that the fidelity has decreased dramatically, in comparison to the other two situations, to a value of 0.76 when compared to $\rho_2^\mathrm f$. However, this can be improved by using shorter pulses, $\sigma_1=\sigma_2=\SI{0.15}{ms}$, achieving a fidelity of 0.82, shown in Figure \ref{stirap(0+1)_1K}, found using the analysis of the parameter space in Appendix \ref{Dependence and Optimization of Parameters}. This demonstrates that temperature must be balanced with the adiabaticity of the system to achieve optimal state transfer. 
\begin{figure}[t]
    \centering
    \includegraphics[width=0.99\linewidth]{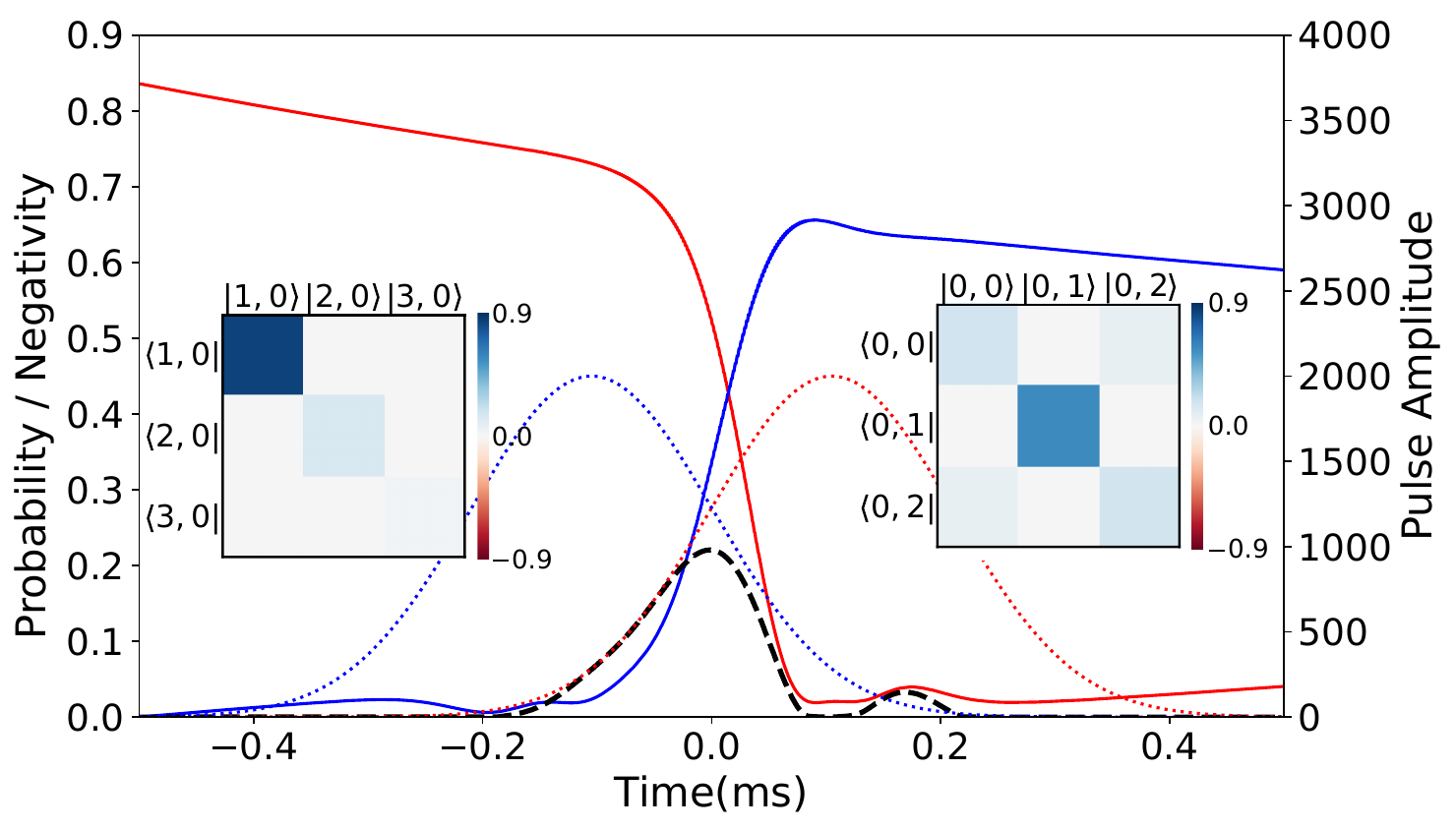}
    \caption{Time evolution of STIRAP of the state $\rho^i_1$ at \SI{1}{K}. Solid: Single-phonon probability amplitude, $n_1$ (red) and $n_2$ (blue) for the first and second resonators, respectively. Dotted: Gaussian profiles, indicating the STIRAP Stokes (red) and pump (blue) pulse sequence. Inset Hinton plots are included for \SI{-0.5}{ms} (left) and \SI{0.5}{ms} (right), representing the density matrix at the start and end of STIRAP, with the color mapping (blue for positive, red for negative) indicating the phase of the off-diagonal elements. Black Dashed: Negativity $\mathcal{N}(\rho_{12})$, reaching a peak during the pulse overlap at \SI{0}{ms}. The final state transfer fidelity is 0.82 when measured against the state $\rho^f_2$.}
    \label{stirap(0+1)_1K}
\end{figure}

Additionally, how $\mathcal{N}(\rho_{12})$ changes at \SI{1}{K} can be determined, shown by the black dashed line in Figure \ref{stirap(0+1)_1K}. The negativity increases during the STIRAP process, peaking at $\mathcal{N} \approx 0.22$. Despite the significantly higher thermal occupancy, the appearance of a non-zero negativity confirms that quantum entanglement is still established between the mechanical resonators during the pulse overlap. The rapid decay of $\mathcal{N}(\rho_{12})$ back to zero immediately following the pulses shows the effects of decoherence from the $1\text{ K}$ environment. In this regime, the rate of thermal phonon scattering is sufficient to quickly “wash out” the off-diagonal coherence in the partially transposed density matrix $\rho_{12}^{\Gamma}$ \cite{Negativity_entangled}, driving the mechanical system toward a purely classical, separable state even before the population transfer is fully finalized.

The results obtained by examining these three cases underscore the fundamental trade-offs inherent in optomechanical STIRAP. As such, it is necessary to write down the requirements for high-fidelity state transfer. First, the resolved sideband regime ($\omega_i \gg \kappa$) should be satisified to ensure coherent interaction. For effective spectral selectivity and to prevent mode cross-talk, the mechanical mode splitting must be sufficiently large ($|\omega_1 - \omega_2| \gg \kappa$). Furthermore, the adiabaticity condition requires that $\tau_\mathrm{transfer} \cdot \max(G_i^2) \gg \kappa$ to maintain the dark state throughout the process. Finally, to protect the quantum state from thermalization, thermal effect must be slow on the time scale of the transfer ($\frac{\hbar Q}{kT} \gg \tau_\mathrm{transfer}$). Together, the adiabaticity condition  and the weak thermalization requirement result in the inequality: 
\begin{equation}
\label{ineq}
\frac{\kappa}{\mathrm{max}(G_i^2)}\ll \tau_\mathrm{transfer}\ll \frac{\hbar Q}{kT}
\end{equation}
The transfer time, balancing these constraints, is given by $\tau_\mathrm{transfer} = \sqrt{\frac{\kappa}{\max(G_i^2)} \frac{\hbar Q}{kT}}$. Utilizing the parameters from Table~\ref{tab:stirap_parameters} and $T = \SI{10}{mK}$ yields $0.004 \ll 1$, while at $T = \SI{1}{K}$ the ratio is $0.04 \ll 1$, both of which are well within a regime of high state transfer fidelities. 




\subsection{Fractional STIRAP}
In this section, fSTIRAP is investigated using the parameters found in Table \ref{tab:stirap_parameters} and optical pulses from \eqref{fractional pumps}. It can be noted that to achieve a high degree of adiabaticity, $\tau_f=\sigma /1.25$, found using the analysis of the parameter space in Appendix \ref{Dependence and Optimization of Parameters}. The same three cases considered in Section~\ref{STIRAP}, applied to fSTIRAP, are studied in the following.

\begin{figure}[!htb]
    \centering
    \includegraphics[width=0.99\linewidth]{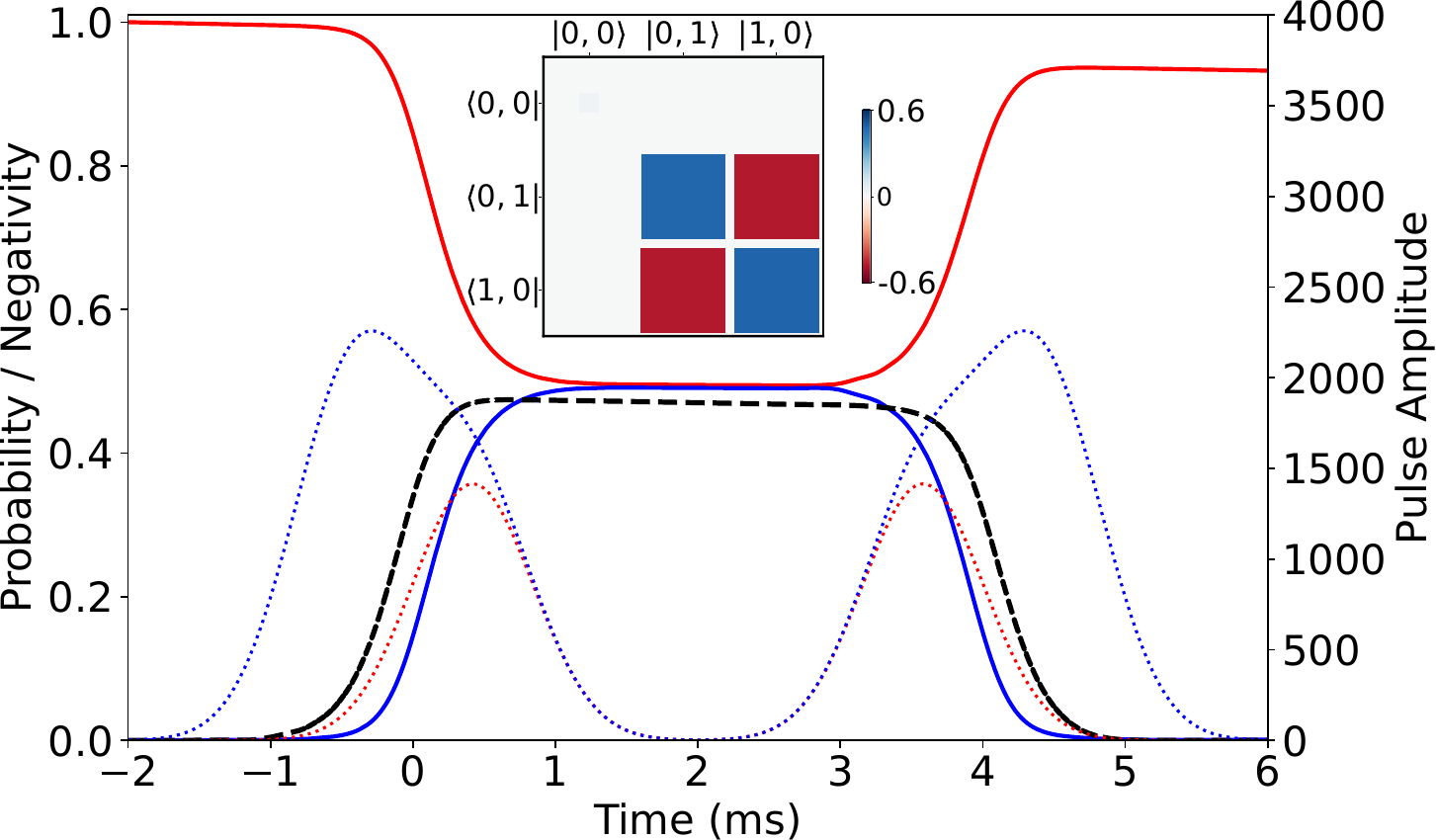}
    \caption{Time-evolution of the single phonon probability amplitude of a single phonon Fock state during fractional STIRAP and, reverse fractional STIRAP at \SI{10}{mK}. At time 2 ms the pulses are off and the system is close to the Bell state $\ket{\Psi^-}$. Red Solid: Mechanical mode 1. Blue Solid: Mechanical mode 2. Red Dotted: Stokes pulse. Blue Dotted: Pump pulse. Inset Hinton plot displays the real part of the density matrix $\rho_{12}$ at \SI{2.0}{ms}, with red and blue regions indicating positive and negative values that characterize the coherence of the evolved state. Black Dashed: Negativity, $\mathcal{N}(\rho_{12})$. The negativity reaches a stable plateau of $\mathcal{N} \approx 0.48$ between \SI{0}{ms} and \SI{4}{ms}. The fidelity of this process against a single phonon Fock state is 0.971.}
    \label{fig:fSTIRAP_comparison}
\end{figure}

First, fSTIRAP of a single phonon Fock state is simulated, which can be found in Figure~\ref{fig:fSTIRAP_comparison}, where a single phonon is placed into a superposition between the two mechanical modes at \SI{10}{mK}. The fSTIRAP pulses from \eqref{fractional pumps} are shown (dotted) in Figure~\ref{fig:fSTIRAP_comparison} from \SI{-2.0}{ms} to \SI{2.0}{ms}.

Additionally, this process can be reversed by applying a time reversed sequence of pulses,
\begin{equation}
\label{rfractional pumps}
\begin{split}
    \alpha_1(t) &= \alpha_0 \sin(\theta) e^{{-(-t - \tau_f)}^2/\sigma^2}\\
    \alpha_2(t) &= \alpha_0 e^{{-(-t + \tau_f)}^2/\sigma^2} + {\alpha_0 \cos(\theta) e^{-{(-t - \tau_f)}^2/\sigma^2}}
\end{split}
\end{equation}
which produces the right side of Figure~\ref{fig:fSTIRAP_comparison} and can be seen from \SI{2.0}{ms} to \SI{6.0}{ms} (dotted). This reversal returns a single phonon transferred by a fractional STIRAP scheme, into the original single phonon Fock state with a fidelity of 0.971 at \SI{10}{mK}. The reduced density matrix $\rho_{12}$, traced over the cavity degrees of freedom, is graphically represented by a Hinton plot in Figure~\ref{fig:fSTIRAP_comparison}, where it can be seen that negative (red) values appear in the off-diagonal elements and the state closely matches with the Bell state, $\ket{\Psi^-}=\frac{1}{\sqrt{2}}(\ket{0}_1\ket{1}_2-\ket{1}_1\ket{0}_2)$, indicated by the high degree of fidelity.

The entanglement generated during fSTIRAP and its subsequent reversal can be observed through $\mathcal{N}(\rho_{12})$, shown as the black dashed line in Figure~\ref{fig:fSTIRAP_comparison}. As the initial fSTIRAP pulses drive the system into $\ket{\Psi^-}$, the negativity rises to a peak value of $\mathcal{N} \approx 0.48$. Upon the application of the time-reversed sequence \eqref{rfractional pumps}, $\mathcal{N}(\rho_{12})$ follows a symmetric fall back to zero. 
The preservation of the negativity's profile throughout the \SI{4}{ms} hold period  underscores the high fidelity ($0.971$) of the process at \SI{10}{mK}.

\begin{figure}[!htb]
    \centering
    \includegraphics[width=0.50\linewidth]{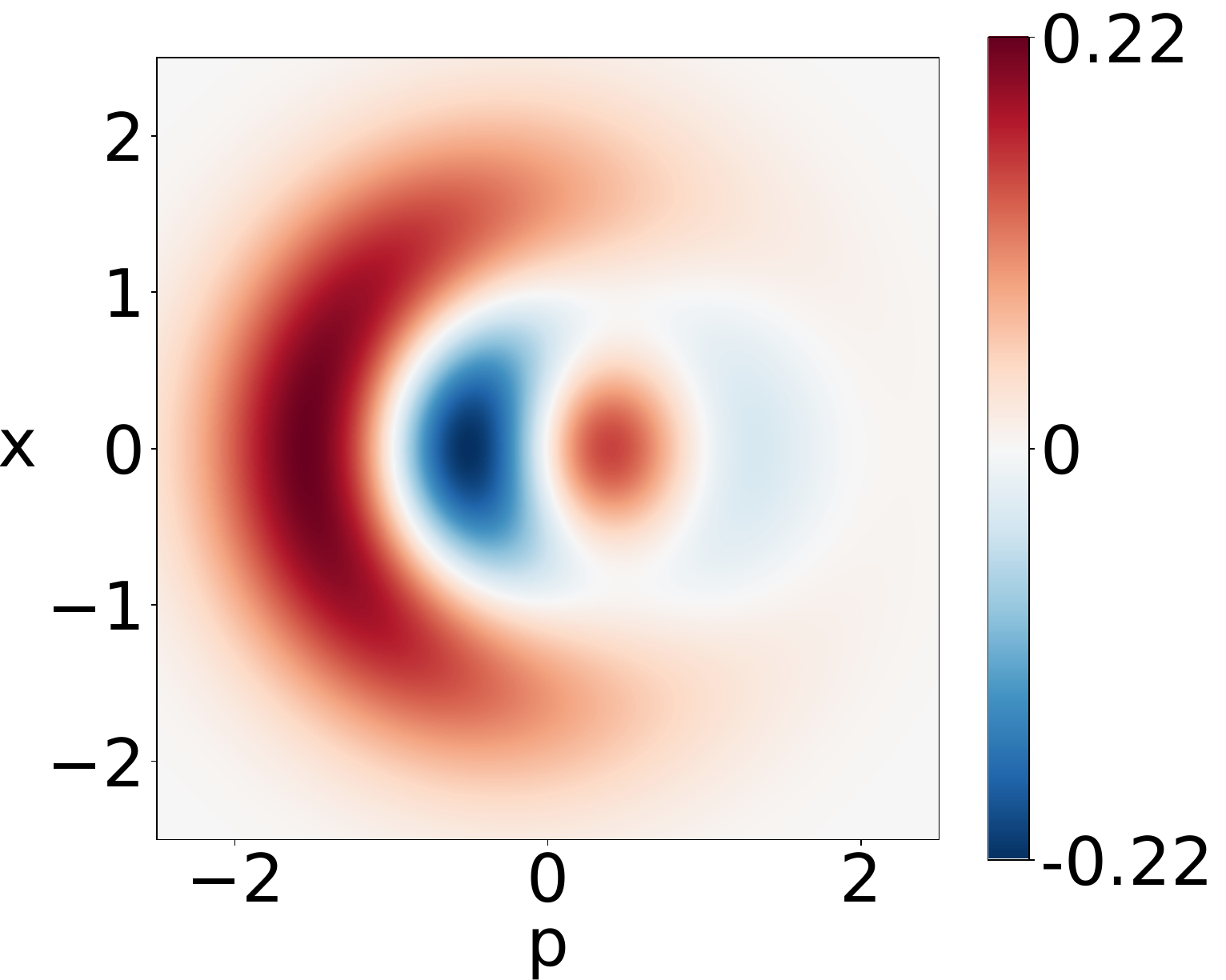}
    \caption{Wigner quasi-probability distribution $W(x,p)$ for the density matrix $\rho_{12}$ at \SI{2.0}{ms} in Figure~\ref{fig:fSTIRAP_comparison}. Here, $x$ and $p$ represent the dimensionless phase space coordinates corresponding to the mechanical modes' position and momentum relative to the zero-point fluctuations. Red and blue regions indicate positive and negative values, respectively, with the negative components characterizing the non-classical nature of the state, which has fidelity of 0.98 when compared to $\ket{\psi^-}$ Bell state.}
    \label{wigner_fstirap}
\end{figure}

The behavior of the density matrix in Figure~\ref{fig:fSTIRAP_comparison} associated with the mechanical modes at \SI{2.0}{ms}, can be compared to that of a standard Bell state, through the use of the Wigner quasi-probability distribution, $W(x,p)=\frac{1}{\pi}\int_{-\infty}^{\infty}dy\bra{x+y}\rho\ket{x-y}e^{-ipy}$, where $x$ and $p$ are the dimensionless phase-space variables. These correspond to the dimensionless operators $\hat{X}=\hat{q}\sqrt{\frac{m\omega}{\hbar}}$ and $\hat{P}=\hat{p}\sqrt{\frac{1}{m\omega\hbar}}$, which are dimensionless position and momentum operators of the two modes' mechanics relative to the zero-point fluctuations of the mechanical modes $x_{\mathrm{zpf},i}=\sqrt{\frac{\hbar}{2m\omega_i}}$ \cite{Wigner-optomechanics}\cite{Clarke_2018}. The state at \SI{2.0}{ms} closely matches with that of $\ket{\Psi^-}$, with fidelity 0.98, and matches with the phase plot of the specified Bell state, as seen in Figure~\ref{wigner_fstirap}. Deviations from the ideal Bell state $\ket{\Psi^-}$ are attributed to thermal effects, cavity leakage, and non-ideal adiabaticity. This implies that fSTIRAP yields a route to generate maximally entangled states.

Next, the case where the temperature is a more realistic value, \SI{50}{mK}, is considered, and fSTIRAP is performed using the state found in \eqref{fstirap_real_initial} as the initial state. The fidelity of this transfer is 0.87 when compared to the Bell state $\ket{\Psi^-}$, which is predominantly due to the higher degree of thermalization, non-single phonon population of the initial state (due to experimental limitations in state preparation \cite{Laser_Cooling}) and low pumping power (limited by the cavity linewidth and heating effects). The first and third of these factors must be balanced, as if the resonator is pumped too hard, thermal phonons are introduced into the mechanical mode via optical heating; if the mechanical modes are pumped for too long, thermalization will introduce phonons from the environment, but if the system is pumped harder or longer, a higher degree of state transfer can be achieved.

Finally, the behavior of fSTIRAP and reverse fSTIRAP at \SI{1}{K} is simulated, to emphasize the need for state-of-the-art cryogenic cooling and the sensitivity of phase terms to temperature. 
At this temperature and time scale, the fidelity of the target state being $\ket{\Psi^-}$ after fSTIRAP is only 0.60. 

\begin{figure}[!htb]
    \centering
        \includegraphics[width=0.99\linewidth]{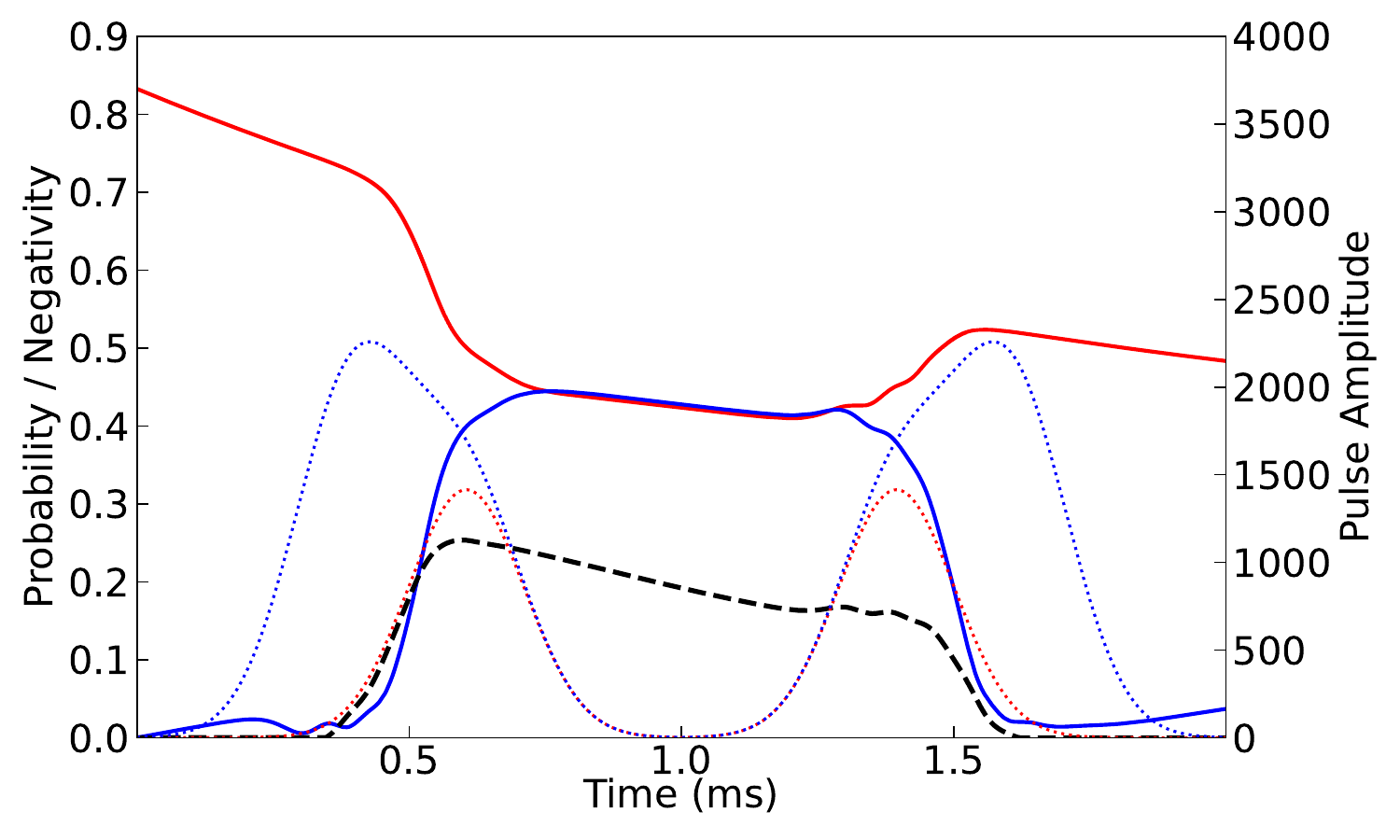}
    \caption{Time evolution of the single phonon population of mode 1 (red solid) and mode 2 (blue solid) during fSTIRAP at \SI{1}{K} of the states $\rho^\mathrm i_1$ and $\rho^\mathrm b_2$. Dotted red and blue lines show the fSTIRAP pulse shapes, with pulse widths of $\sigma_1 = \sigma_2 = \SI{0.15}{ms}$. Black Dashed: Negativity $\mathcal{N}(\rho_{12})$ with a maximum at $\sim$\SI{0.6}{ms} of $\mathcal{N}(\rho_{12})\approx 0.25$ followed by a steady decay. The state transfer has a fidelity of 0.77, compared to the state $\ket{\Psi^-}$ at \SI{1}{ms}.}
    \label{1K_bellstate.pdf}
\end{figure}

The state transfer could potentially be improved by reducing the timescale of the experiment (becoming less adiabatic) at this temperature. This can be done by reducing the pulse widths associated with \eqref{fractional pumps}. By using $\sigma_1=\sigma_2=\SI{0.15}{ms}$, found using the analysis in Appendix \ref{Dependence and Optimization of Parameters}, the result in Figure~\ref{1K_bellstate.pdf} can be obtained, which has a fidelity of 0.77 when compared to $\ket{\Psi^-}$ at \SI{1}{ms}. However, it can be seen from the behavior of the single phonon probability amplitude (solid) for mechanical mode 1 (red) and 2 (blue) that fSTIRAP has become less adiabatic.

Furthermore, how the entanglement is affected by increased thermal noise can be observed, by examining $\mathcal{N}(\rho_{12})$, indicated by the black dashed line in Figure \ref{1K_bellstate.pdf}. At \SI{1}{K}, the peak negativity reaches a value of $\mathcal{N} \approx 0.25$, during the initial pulse overlap. Unlike the stable negativity observed at 
\SI{10}{mK} in Figure \ref{fig:fSTIRAP_comparison}, the negativity in Figure \ref{1K_bellstate.pdf} shows a continuous decay from \SI{0.5}{ms} to \SI{1.5}{ms}. This profile highlights the sensitivity of quantum coherence to the thermal environment. 

By comparing these three cases, it can be determined that the temperature of the environment is a critical parameter in the context of the creation and manipulation of a mechanical Bell state, once the adiabatic parameters have been satisfied. The results of these cases for STIRAP and fSTIRAP can be found in Table~\ref{tab:stirap_temp}.
\begin{table}[htbp]
    \centering
    \caption{STIRAP and fSTIRAP Fidelity Comparison. $\rho_1^i$ is the density matrix shown in \eqref{fstirap_real_initial}. $\rho_2^f$ is a density matrix with the same probability amplitudes as $\rho_1^i$, but in mechanical mode 2. $\ket{\Psi^-}$ is the Bell state $\ket{\Psi^-}=\frac{1}{\sqrt{2}}(\ket{0}_1\ket{1}_2-\ket{1}_1\ket{0}_2)$.}
    \label{tab:stirap_temp}
    \addtolength{\tabcolsep}{-3pt} 
    \begin{tabular*}{\columnwidth}{@{\extracolsep{\fill}}lllll}
        \toprule
        \textbf{Mode} & \textbf{Temp.} & \textbf{Initial State} & \textbf{Target} & \textbf{Fidelity} \\
        \midrule
        STIRAP & \SI{10}{mK} & $\frac{1}{\sqrt{2}}(\ket{0}_1+\ket{1}_1)$ & $\frac{1}{\sqrt{2}}(\ket{0}_2-\ket{1}_2)$ & 0.98 \\
        STIRAP & \SI{50}{mK} & $\rho_{1}^\mathrm{i}$ & $\rho_2^\mathrm{f}$ & 0.93 \\
        STIRAP & \SI{1}{K}   & $\rho_{1}^\mathrm{i}$ & $\rho_2^\mathrm{f}$ & 0.82 \\
        \midrule
        fSTIRAP & \SI{10}{mK} & $\ket{1}_1$ & $\ket{\Psi^-}$ & 0.98 \\
        fSTIRAP & \SI{50}{mK} & $\rho_{1}^\mathrm{i}$ & $\ket{\Psi^-}$ & 0.87 \\
        fSTIRAP & \SI{1}{K}   & $\rho_{1}^\mathrm{i}$ & $\ket{\Psi^-}$ & 0.77 \\
        \bottomrule
    \end{tabular*}
\end{table}
\subsection{Entanglement Verification}

In addition to developing a scheme to read out the state after transfer, found in Appendix~\ref{experimental appendix}, it would be useful to perform a verification of the entanglement produced by fSTIRAP.  There are several ways one might be able to do this, and we will provide an example using STIRAP pulse sequences applied to two different states. 

\begin{figure}[!ht]
\centering
\includegraphics[width=0.99\linewidth]{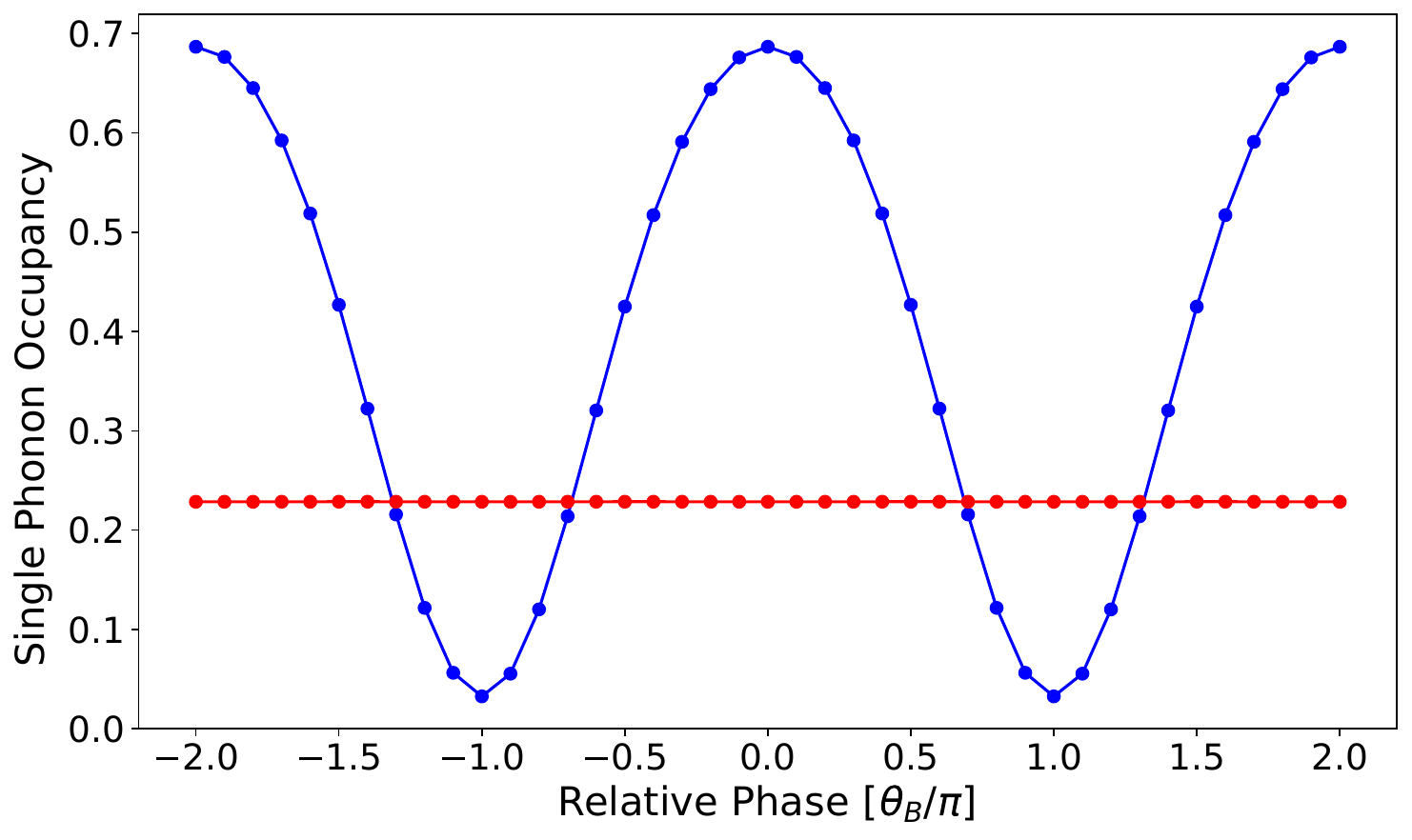}
\caption{Single phonon occupation of the target state as a function of the relative phase $\theta_B/\pi$ of the reversed fSTIRAP pulse sequence. The blue curve tracks the occupancy for the initial state $\rho_1^i$ with the first fSTIRAP pulse phase fixed at $\theta_A = 0$, exhibiting clear interference fringes with maxima at $\theta_B = 0, \pm 2\pi$ and minima occurring at $\theta_B = \pm \pi$. The red horizontal line represents the occupation for a thermal state ($\bar{n}_1 = 0.5$) transferred under the same conditions. These results are obtained at \SI{50}{mK} using the experimental parameters detailed in Table~\ref{tab:stirap_parameters}.}  
\label{rfstirapvphase.pdf}
\end{figure}

First, it is assumed that after fSTIRAP, the mean occupancy of mechanical modes are identical $\bar{n}_1=\bar{n}_2=\bar{n}$, and the density matrix takes the form of a thermal state, $\rho_\mathrm{th}=\sum_{n_1,n_2} P_{n_1}P_{n_2}\ket{0,n_1,n_2}\bra{0,n_1,n_2}$. By applying a  unitary evolution associated with reverse fSTIRAP $\hat U_\mathrm{rs}(\phi_2)$, where $\phi_2$ is the relative phase between the pump and Stokes pulses, it can be observed how the number operator of mechanical mode 1 in the Heisenberg picture at the end of the scheme is applied,
\begin{equation}
\begin{gathered}
\label{rfstirap transformation}
\hat{n}_1= \hat U^\dagger_\mathrm{rs}\hat{b}^\dagger_1\hat{b}_1\hat U_\mathrm{rs},
    \\ \hat{n}_1= \frac{1}{2}\hat{b}^\dagger_1\hat{b}_1+\frac{1}{2}\hat{b}^\dagger_2\hat{b}_2+\frac{1}{2}(e^{-i\phi_2}\hat{b}^\dagger_1\hat{b}_2+e^{i\phi_2}\hat{b}^\dagger_2\hat{b}_1).
\end{gathered}
\end{equation}
Taking the expectation value of $\hat{n}_1$, $\braket{\hat{n}_1}=\frac{1}{2}\bar{n}_1+\frac{1}{2}\bar{n}_2$, where the ``beam-splitter'' terms have disappeared, as there exist no off-diagonal terms in $\rho_\mathrm{th}$. This means that the result of reverse fSTIRAP for a product state will be constant. The result of this is plotted in Figure~\ref{rfstirapvphase.pdf} in red, under realistic parameters discussed in the previous section. It holds a single phonon probability amplitude of  $\sim0.22$, set by $\bar{n}_1=\bar{n}_2=0.5$ in the state after fSTIRAP. The result in Figure~\ref{rfstirapvphase.pdf} is lower than the ideal due to dissipation included in the simulation.

Next fSTIRAP is considered, using the states found in \eqref{fstirap_real_initial}, and applying the same unitary evolution using \eqref{rfstirap transformation}, yields $\braket{\hat{n}_1}=\frac{1}{2}+\frac{1}{2}(e^{-i\phi_2}\frac{1}{2}e^{-i\phi_1}+e^{i\phi_2}\frac{1}{2}e^{i\phi_1})$. This can be rewritten as $\braket{\hat{n}_1}=\frac{1}{2}(1+V\cos(\phi_1-\phi_2))$, where $V$ is written to include sources of dissipation and decoherence. In terms of the sources of dissipation and decoherence present within the simulation, including the detector efficiency $\eta_\mathrm d$ and the readout efficiency $\eta_\mathrm r$, $V$ will take the form 
\begin{equation}
    V=\eta_\mathrm d\eta_\mathrm r \exp\!\left(-\left(\frac{\kappa+\Gamma_\mathrm m}{2}+\Gamma_\mathrm m \bar{n}_\mathrm b\right)\tau_\mathrm{w}\right),
\end{equation}
where $\tau_w$ is the time separation between the pulses for fSTIRAP and reverse fSTIRAP. The result of this applied to the system, at \SI{50}{mK} and the parameters found in Table~\ref{tab:stirap_parameters}, can be seen in Figure~\ref{rfstirapvphase.pdf} (blue) where the single phonon probabilities are plotted as $\phi_2$ is changed for a fixed $\phi_1=0$, under ideal detection and readout.

In the case presented here, the state can be read out using a red-detuned pulse of $\omega_{\mathrm L1}$ of \SI{73.2}{\micro s} with 99\% efficiency. Upon readout, the probability of detection of a single photon at a detector would go as $P_{1}(\phi_1,\phi_2)\sim n_{1}(\phi_1,\phi_2)$. The inefficiencies $\eta_\mathrm d$ and $\eta_\mathrm r$ would apply a linear scaling to the result in Figure~\ref{rfstirapvphase.pdf}, but would not affect the signature of entanglement, as the visibility of the interferometric fringe can give the degree of entanglement exhibited by the system such that a maximally entangled state will have a visibility of $V=1$. If damping coefficients of $\eta_\mathrm d = 0.075$ and $\eta_\mathrm r = 0.99$ are applied from Appendix \ref{experimental appendix}, we would see that the single phonon occupation (blue) in Figure \ref{rfstirapvphase.pdf} would have an additional attenuation and would achieve a peak value of 0.0525, showing that a main experimental hurdle is the detection efficiency.

\section{Conclusion}
We presented a quantum-mechanical treatment of optomechanical STIRAP and fractional STIRAP (fSTIRAP). By utilizing a collective mode analysis, we demonstrated that the state transfer is mediated by a dark state that decouples the resonators from the lossy optical cavity. We analytically showed that the transfer of an $n$-phonon Fock state can be mapped onto a multi-state chain of $2n+1$ states, where the adiabatic evolution preserves the population in the dark state. A key analytical finding is of a parity-dependent phase shift: the relative phase of the generated superposition state is 
determined by the parity of the initial phonon number states involved. Our analytical study also highlighted a clear distinction between quantum and classical-like initial states. We showed that fSTIRAP acting on a single-phonon Fock state generates a maximally entangled Bell state, and that fSTIRAP acting on a coherent state results in a product state, regardless of the mixing angle $\theta$. To validate these findings under realistic conditions, we performed numerical simulations using a master equation approach and confirmed the creation of an mechanical Bell state through the use of Wigner probability distributions. These simulations quantify the impact of thermal decoherence on the state fidelity and the negativity $\mathcal{N}(\rho_{12})$. Our results indicate that while entanglement can be established at temperatures up to 1 K, high-fidelity state preparation requires state-of-the-art cryogenic and optical cooling near 10 mK to prevent thermal phonons from washing out the non-classical correlations. Finally, we proposed an interferometric method to quantify mechanical entanglement by applying a time-reversed fSTIRAP pulse sequence, providing an experimental path to verify coherence and entanglement without full state tomography.

\begin{acknowledgments}
We gratefully acknowledge support via the UC Santa Barbara NSF Quantum Foundry funded via the Q-AMASE-i program under award DMR-1906325, support from NSF Award No. 2427169, 2137740 and Q-AMASE-i, through Grant No. DMR-1906325, and from NWO Quantum Software Consortium (Grant No. 024.003.037). We would also like to acknowledge support from the Eddleman Center for Quantum Innovation.
\end{acknowledgments}

\appendix

\section{Energy gap of the dark state}
\label{energy gap dark state}

In the following, the energy gap of the dark state is considered, and is determined by first taking \eqref{hamiltonian rwa} as the definition of the Hamiltonian $\hat H$.
Define an angle $\tan\vartheta=\Omega_1/\Omega_2$, where $\Omega_i\coloneqq2G_{i,i}$.
Define:
\begin{equation}
\begin{split}
\hat b_+&=\hat b_1e^{i\phi_1}\sin\vartheta+\hat b_2e^{i\phi_2}\cos\vartheta,\\
\hat b_-&=\hat b_1e^{i\phi_1}\cos\vartheta-\hat b_2e^{i\phi_2}\sin\vartheta.
\end{split}
\end{equation}
These definitions agree with \eqref{b-}, and they agree with \eqref{collective without rwa}
in the case of $\phi_i=0$.
One can verify the canonical commutation relations
\begin{equation}
\left[\hat b_\pm,\hat b_\pm^\dagger\right]=1,\quad
\left[\hat b_\pm,\hat b_\pm\right]
=\left[\hat b_\pm,\hat b_\mp\right]=0
\end{equation}
to see that they are annihilation operators of independent boson modes.
Physically, they represent collective modes involving both mechanical modes.
Operators $\hat a^\dagger\hat a$, $\hat b_+^\dagger\hat b_+$, and $\hat b_-^\dagger\hat b_-$
form a complete set of commuting operators, so their eigenvectors form an orthonormal basis
whose each basis vector $\left|n_\mathrm c,n_+,n_-\right>$
(not to be confused with $\left|n_\mathrm c,n_1,n_2\right>$) is labeled
by the eigenvalues $n_\mathrm c$, $n_+$, and $n_-$.
The numbers $n_\mathrm c,n_+,n_-$ can take any nonnegative integers.
An identity that will be useful later is
\begin{equation}
\label{partial theta}
\partial_\vartheta\left|n_\mathrm c,n_+,n_-\right>
=\left(\hat b_+\hat b_-^\dagger-\hat b_-\hat b_+^\dagger\right)\left|n_\mathrm c,n_+,n_-\right>.
\end{equation}
One can easily prove this by noticing that $\partial_\vartheta\hat b_+=\hat b_-$ and $\partial_\vartheta\hat b_-=-\hat b_+$.

Using Schwinger's definition of angular momentum \cite{Schwinger1952},
we define
\begin{equation}
\begin{split}
\hat J_x&=\frac12\left(\hat a^\dagger\hat b_++\hat a\hat b_+^\dagger\right),\\
\hat J_y&=\frac i2\left(\hat a^\dagger\hat b_+-\hat a\hat b_+^\dagger\right),\\
\hat J_z&=\frac12\left(\hat b_+^\dagger\hat b_+-\hat a^\dagger\hat a\right).
\end{split}
\end{equation}
One can verify that $\left[\hat J_j,\hat J_k\right]=i\epsilon_{jkl}\hat J_l$ to see that they are angular momentum operators.
Operators $\hat J^2=\hat J_x^2+\hat J_y^2+\hat J_z^2$, $\hat J_z$, and $\hat b_-^\dagger\hat b_-$ form a complete set of commuting operators,
so their eigenvectors form an orthonormal basis
whose each basis vector $\left|j,m_z,n_-\right>$
is labeled by the eigenvalues $j\left(j+1\right)$, $m_z$, and $n_-$.
The numbers $j,n_-$ can take any nonnegative integers,
and $m_z=-j,\ldots,j$.
Similarly, $\hat J^2$, $\hat J_x$, and $\hat b_-^\dagger$ form a complete set of commuting operators,
and the basis eigenvectors are labeled by $j$, $m_x$, and $n_-$.

From basic quantum mechanics, it is known that $\left|j,m_x,n_-\right>$
is a linear combination of states $\left|j,m_z,n_-\right>$ with different values of $m_z$.
Therefore, noticing \eqref{partial theta} and
\begin{equation}
\left|j,m_z,n_-\right>=\left|n_\mathrm c=j-m_z,n_+=j+m_z,n_-\right>,
\end{equation}
we know that $\partial_\vartheta\left|j,m_x,n_-\right>$ is a linear combination of
states $\left|j\pm1/2,m_x',n_-\right>$ with different values of $m_x'$.
Notice that $m_x'$ must differ $m_x$ by a half-integer but not an integer.

Using angular momentum operators, the Hamiltonian can be rewritten as
\begin{equation}
\hat H=\Omega\hat J_x,
\end{equation}
where $\Omega\coloneqq\sqrt{\Omega_1^2+\Omega_2^2}$.
Therefore, every basis vector $\left|j,m_x,n_-\right>$ is an eigenstate with energy $E_{j,m_x,n_-}=m_x\Omega$,
which is time-independent.

For abbreviation, denote a tuple $A=\left(j,m_x,n_-\right)$.
Suppose that $\sum_Ac_Ae^{-itE_A}\left|A\right>$ is a solution of the Schr\"odinger equation.
Then, from the Schr\"odinger equation,
\begin{equation}
\label{coefficient eom}
\dot c_A=\dot\vartheta\sum_{A'}c_{A'}e^{it\left(E_A-E_{A'}\right)}\left<A\right|\partial_\vartheta\left|A'\right>.
\end{equation}
In a standard proof of the adiabatic theorem, one will work with the assumption that $E_A$ is nondegenerate and gaped,
so the factor $e^{it\left(E_A-E_{A'}\right)}$ is always fast oscillating in the adiabatic limit except when $A=A'$.
Our situation here is different from the assumption of the standard adiabatic theorem in that
every energy level is degenerate.
However, every term in the right-hand side of \eqref{coefficient eom} is exactly zero
whenever $E_A-E_{A'}=0$
because $\left<A\right|\partial_\vartheta\left|A'\right>$ can only be nonzero when $m_x'$ differs from $m_x$ by a half-integer but not an integer.
Therefore, different eigenstates do not mix with each other during an adiabatic passage even if they have the same energy.
The energy gap that a nonadiabatic transfer of population must overcome is $\Omega/2$,
the smallest value of $\left|E_A-E_{A'}\right|$ such that $\left<A\right|\partial_\vartheta\left|A'\right>\ne0$.

\section{Adiabaticity condition}
\label{section adiabaticity condition}

In this section, we follow \cite{fSTIRAP} to express the adiabaticity condition
for the scheme described by \eqref{fractional pumps} for an optomechanical system.

For appreciable non-adiabatic transitions to occur, two conditions must be satisfied:
\begin{equation}
    \left|\dot\vartheta(t)\right|\gtrsim\frac12\Omega(t),\quad
\left|\dot\vartheta(t)\right|\gtrsim\frac1\sigma.
\end{equation}
Here, $\Omega(t)=\sqrt{\Omega_1(t)^2+\Omega_2(t)^2}$ is twice the energy gap of the dark state
(see Appendix~\ref{energy gap dark state} for a justification of this claim).
Non-adiabatic transitions are most likely to occur when $\dot\vartheta(t)$ is maximized, which is  when $t=0$.
At that time,
\begin{equation}
\begin{split}
\dot\vartheta(0) = \dot\vartheta_\mathrm{max} &= \frac{2\tau}{\sigma^2}\tan\frac\theta2, \\
\Omega(0) &= 2\Omega_0 e^{-\tau^2/\sigma^2} \cos\frac\theta2
\end{split}
\end{equation}
(this is not necessarily when $\Omega(t)$ is maximized).
Because both $\Omega(t)$ and $\dot\vartheta(t)$ are pulse-shaped,
it can also be beneficial to consider their widths $T_{\dot\vartheta}$ and $T_{\Omega}$ (full widths at half maximum),
\begin{equation}
    T_{\dot\vartheta}=\frac{\sigma^2}{\tau}\operatorname{arsinh}\cos\frac\theta2,\quad
T_\Omega\approx 2\tau+2\sigma\sqrt{\ln2}.
\end{equation}

When $\tau$ gets smaller, $\dot\vartheta$ becomes a wider pulse, while $\Omega$ becomes a narrower pulse.
If $\dot\vartheta$ is wider than $\Omega$, non-adiabatic transitions can occur at early and late times,
so $T_{\dot\vartheta}\lesssim T_\Omega$ gives a lower bound on $\tau$. When $\tau$ gets large, $\dot\vartheta(0)$ increases
while $\Omega(0)$ decreases. When $\dot\vartheta$ is a taller pulse than $\Omega$,
non-adiabatic transitions can occur at around $t=0$,
so $\Omega(0)/2\gtrsim n_o\dot\vartheta(0)$ gives an upper bound for $\tau$,
where $n_o$ is a real positively defined arbitrary value that depends on how much non-adiabaticity can be allowed.
Combining the lower bound and the upper bound, an adiabaticity condition for our system can be written
\begin{equation}
\label{adiabaticity condition}
\begin{split}
\sqrt{\ln2+2\operatorname{arsinh}\cos\frac\theta2}-\sqrt{\ln2}\\
\lesssim\frac{2\tau}{\sigma}\lesssim
\sqrt{2W_0\!\left(\frac{\cos^2(\theta/2)}{n_o\sin(\theta/2)}\Omega_0\sigma\right)}.
\end{split}
\end{equation}
where $W_0$ is the Lambert $W$ function.

Using a single phonon optomechanical coupling rate of,  $g_{10}/2\pi=g_{20}/2\pi=\SI{2.5}{Hz}$, a peak amplitude of $\alpha=2000$, a pulse width, $\sigma=\SI{0.6}{ms}$, and setting $n_o=5$. An adiabaticity condition of
$0.29\lesssim\tau/\sigma\lesssim0.89$ for STIRAP ($\theta=\pi/2$) is found. Additionally, a separate condition for fSTIRAP can be written, where $\theta=\pi/4$, $0.35\lesssim\tau/\sigma\lesssim1.18$.

\section{Impact of System Parameters on Transfer Efficiency}
\label{Dependence and Optimization of Parameters}

State transfer efficiency is heavily dependent on the adiabaticity of the state transfer and control parameters $\{\Omega_0, \tau, \sigma_1, \sigma_2, \omega_1, \omega_2\}$. The system is defined by the broader set of parameters in Table \ref{tab:stirap_parameters}, which includes a set of loss parameters associated with an optomechanical system. The simulations in this section are conducted with a single phonon Fock state as the initial state, in the presence of a \SI{10}{mK} thermal bath. This temperature is selected so that the rate of thermalization is small on the time-scale considered.

\begin{figure}[!htb]
        \centering
        \includegraphics[width=0.99\linewidth]{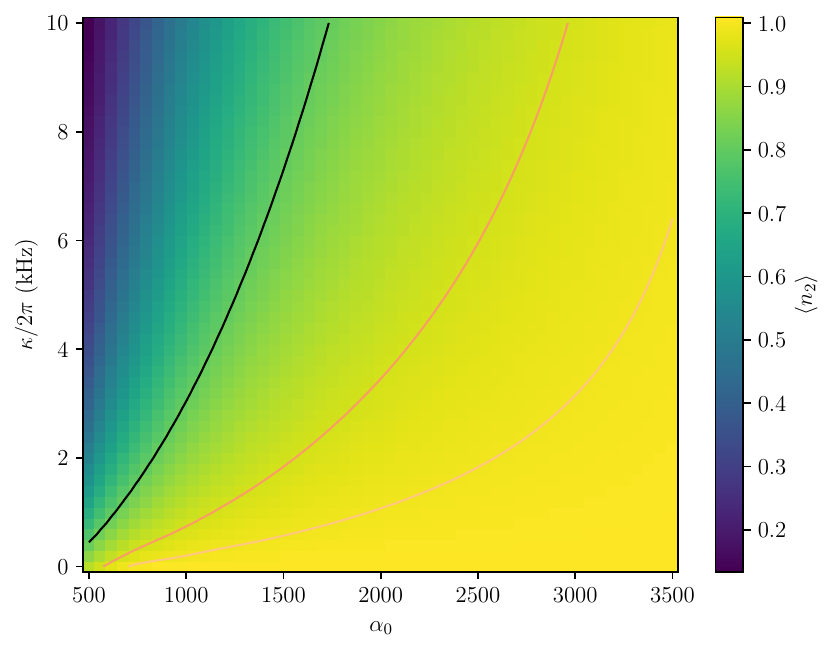}
        \caption{Final state population $\langle n_2 \rangle$ as a function of the cavity linewidth $\kappa/2\pi$ and the peak pulse amplitude $\alpha_0$. The contour lines represent fidelity of 0.80 (black), 0.95 (orange) and 0.99 (pink). The high fidelity region corresponds to the parameter space where the peak optomechanical damping rate $\Gamma_{\text{om}} \approx 4{G_0}^2/\kappa$ is sufficiently large to maintain adiabaticity.}
        \label{fig:amplitude_kappa_sweep}
    \end{figure}

The cavity field strength and cavity linewidths are factors in the efficiency of state transfer. Figure \ref{fig:amplitude_kappa_sweep} illustrates the sensitivity of the final target state population $\langle n_2 \rangle$ on $\kappa$ and $\alpha_0$. When $\kappa$ is large, the system fails to maintain high fidelity state transfer as the cavity induced decoherence outpaces the STIRAP process. Conversely, as $\alpha_0$ increases, the effective coupling strength enhances the adiabaticity of the transfer, placing our system into the dark state. The contour lines indicate the transition boundaries between regions of fidelity of 0.80 (black), 0.95 (orange) and 0.99 (pink).

This behavior is linked to the effective optomechanical coupling $G_0=g_j \alpha_0$. In the resolved sideband regime where $\kappa \ll \omega_m$, consistent with our system parameters, the rate of energy exchange between optical and mechanical modes can be characterized by the optomechanical damping rate, $\Gamma_{\text{om}} \approx 4{G_0}^2/\kappa$ \cite{Optomechanics}. As shown in Figure \ref{fig:amplitude_kappa_sweep}, increasing $\alpha_0$ leads to a higher $\Gamma_{\text{om}}$, which facilitates a faster and more robust transfer to the target phonon state $\langle n_2 \rangle$. However, this enhancement must be balanced against the cavity linewidth $\kappa$; while a larger $\kappa$ can lead to easier coupling, it simultaneously suppresses $\Gamma_{\text{om}}$ for a fixed amplitude, shifting the system out of the high-fidelity (yellow) region.

\begin{figure}[!htb]
    \centering
    \includegraphics[width=0.99\linewidth]{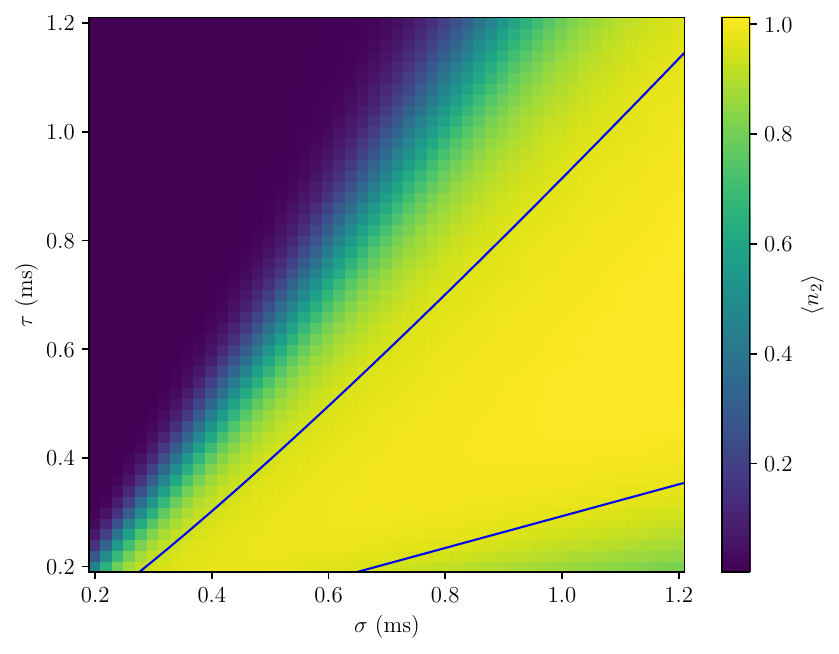}
    \caption{Pulse width ($\sigma$) and pulse separation ($\tau$) versus the expectation value of the number of transferred phonons $\langle n_2\rangle$.
The area between the blue lines is the region that satisfies the adiabaticity condition given in \eqref{adiabaticity condition} for $n=5$, corresponding for a fidelity of state transfer $>0.90$.}
    \label{Time Delay Sweep}
\end{figure}

Sweeping the pulse separation and pulse width versus the expectation value of the final state yields Figure~\ref{Time Delay Sweep}. An optimal region of the state transfer is found, that coincides with the orange contour. It can be clearly seen that there are several regions of behavior within Figure~\ref{Time Delay Sweep}, set by the adiabaticity condition found in \eqref{adiabaticity condition} for $n=5$. That is, when the pulse width becomes narrow with respect to a constant peak amplitude, the adiabaticity of our system is reduced and state transfer becomes non-optimal. Additionally, we can observe that at a fixed value for $\tau$, if $\sigma$ becomes large, then the efficiency of state transfer is reduced. This is due to the pump pulse overlapping with the Stokes pulse, taking us away from the dark state.

\begin{figure}[!htb]
    \centering
    \includegraphics[width=0.99\linewidth]{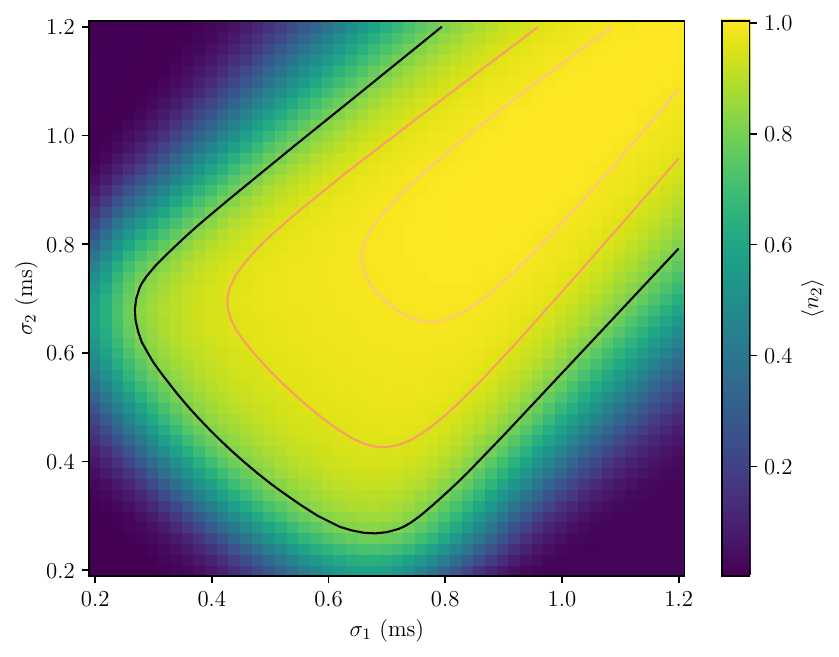}
    \caption{Single phonon STIRAP target state population $\langle n_2\rangle$ plotted against a sweep of  the pulse widths $\sigma_1$ and $\sigma_2$. Optimal values for state transfer are found and indicated by the contour lines as shown. The contours correspond to a fidelity of 0.80 (black), 0.95 (orange) and 0.99 (pink). A decrease in $\langle n_2 \rangle$ is observed at lower pulse widths, where adiabaticity is lost, and at higher pulse widths, where excessive overlap occurs.}
    \label{sigma sweep}
\end{figure}

The pulse widths $\sigma_1$ and $\sigma_2$ can also be varied with respect to one another, to determine their effect on state transfer, as shown in Figure~\ref{sigma sweep}. Additionally, similar behavior can be seen in Figure~\ref{sigma sweep} as in Figure~\ref{Time Delay Sweep}, where if $\sigma_1$ and $\sigma_2$ become small, such that the pulses are no longer adiabatic, state transfer efficiency is reduced. Furthermore, when pulse widths become large, the pulses start to overlap, and the timing of the state transfer protocol overlaps, moving us away from an optimal value.

\begin{figure}[!htb]
    \includegraphics[width=0.99\linewidth]{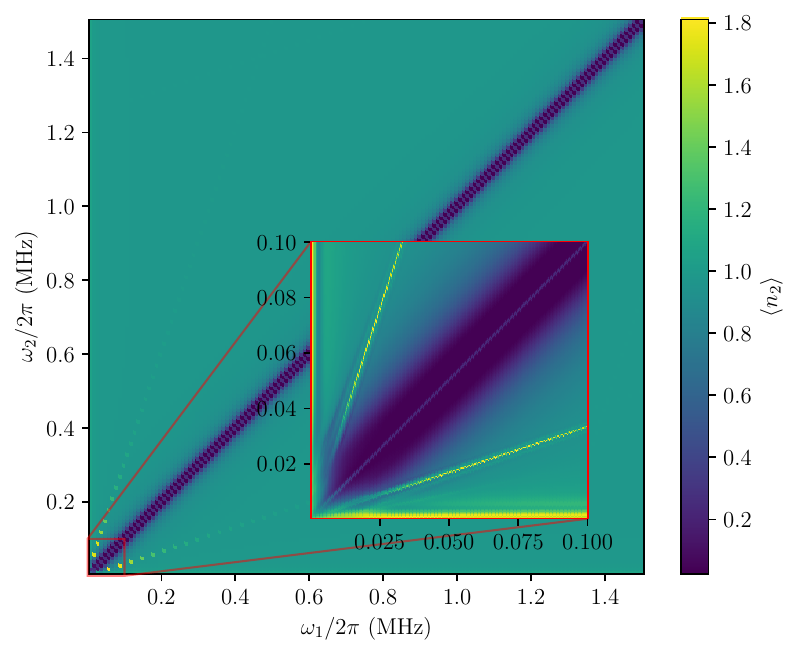}
    \caption{The single phonon STIRAP plotted against a sweep of  the mechanical frequency $\omega_1$ and $\omega_2$. Near $\omega_1=\omega_2$, $\langle n_2\rangle$ is close to zero, but is approximately $1/4$ at $\omega_1=\omega_2$. A zoom in of the behavior near when $\omega_1=\omega_2=0$ shows that when $\omega_1=3\omega_2$ or $\omega_2=3\omega_1$, $\langle n_2\rangle$ is  greater than 1.}
    \label{sweep-omega1-omega2}
\end{figure}

\begin{figure}[!htb]
    \includegraphics[width=0.99\linewidth]{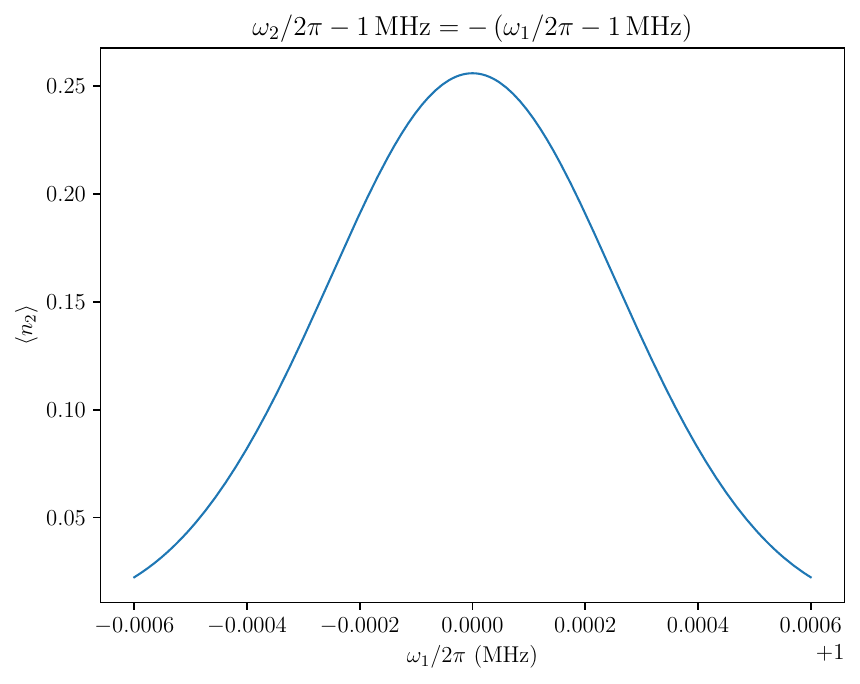}
    \caption{Sensitivity of target state population $\langle n_2 \rangle$ to mechanical frequency detuning. The population is plotted along a segment perpendicular to the degeneracy line $\omega_1 = \omega_2$. The dependency of $\langle n_2\rangle$ on $\omega_1$ is extremely sensitive, as shown.}
    \label{sweep-omega1-omega2-near-diagonal}
\end{figure}

\begin{figure}[!htb]
    \centering
    \includegraphics[width=0.99\linewidth]{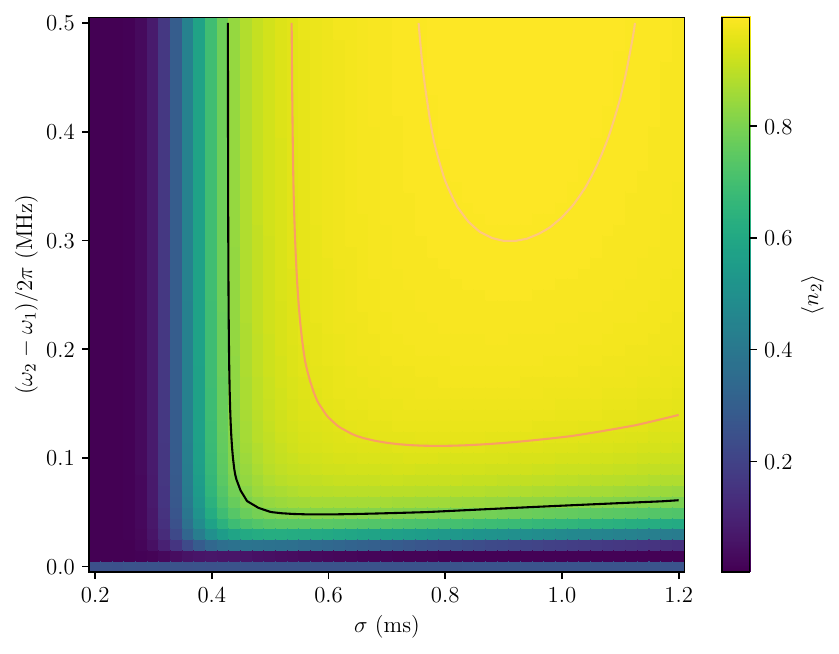}
    \caption{Pulse width, $\sigma$ plotted against mechanical mode frequency difference $\delta$, $\omega_2-\omega_1$. When the pulse width is small the state transfer becomes non-adiabatic and the state transfer becomes highly inefficient. Additionally, when the mechanical mode frequency difference is small the red detuned pulse, is applied to both modes and the initialized state, associated with $\langle n_1\rangle$, is depleted, resulting in low efficiency state transfer. This sets conservative limits for state transfer of $\omega_2-\omega_1 >\SI{0.05}{MHz}$ for  and $\sigma>0.5$ ms. Additionally,  optimal values for state transfer are found and indicated by the contour lines as shown. The contours correspond to a fidelity of 0.80 (black), 0.95 (orange) and 0.99 (pink).}
    \label{sigma-omega sweep}
\end{figure}

The mechanical frequencies $\omega_1$ and $\omega_2$ are swept against each other to determine their effects on state transfer, as shown in Figure~\ref{sweep-omega1-omega2}.The behavior is consistent, except for when $\omega_1-\omega_2$ is close to zero, where the expectation value of the target state reduces to $\langle n_2\rangle=0$. This is caused by energy degeneracy, when $\omega_1=\omega_2$, which causes the red detuned pulse, which is usually first coupled to the target state, $\langle n_2\rangle$, to first become coupled to the initial state $\langle n_1\rangle$. This causes population depletion of the initialized state, reducing the overall population of our system to zero. Additionally, we can plot the mechanical mode frequency difference $\omega_2-\omega_1$ against $\sigma$, shown in Figure~\ref{sigma-omega sweep}, which shows that when the mechanical mode frequencies are close together STIRAP fails. This allows us to set a parameter restriction on the mechanical mode frequency difference of $>\SI{0.05}{MHz}$. In Figure~\ref{sigma-omega sweep}, one can more clearly see the behavior near the line $\omega_1=\omega_2$.
It shows the final value of $\left<n_2\right>$ as $\omega_1$ and $\omega_2$ change along a short segment perpendicular to the line $\omega_1=\omega_2$.
When $\omega_1-\omega_2$ is very close
(much smaller than the scale of it under which the STIRAP fails),
the final $\left<n_2\right>$ is close to $1/4$ instead of being close to $0$, as seen in Figure \ref{sweep-omega1-omega2-near-diagonal}.
It tends towards zero as the difference between $\omega_1$ and $\omega_2$ increases.

\begin{figure}[!htb]
    \centering
    \includegraphics[width=0.99\linewidth]{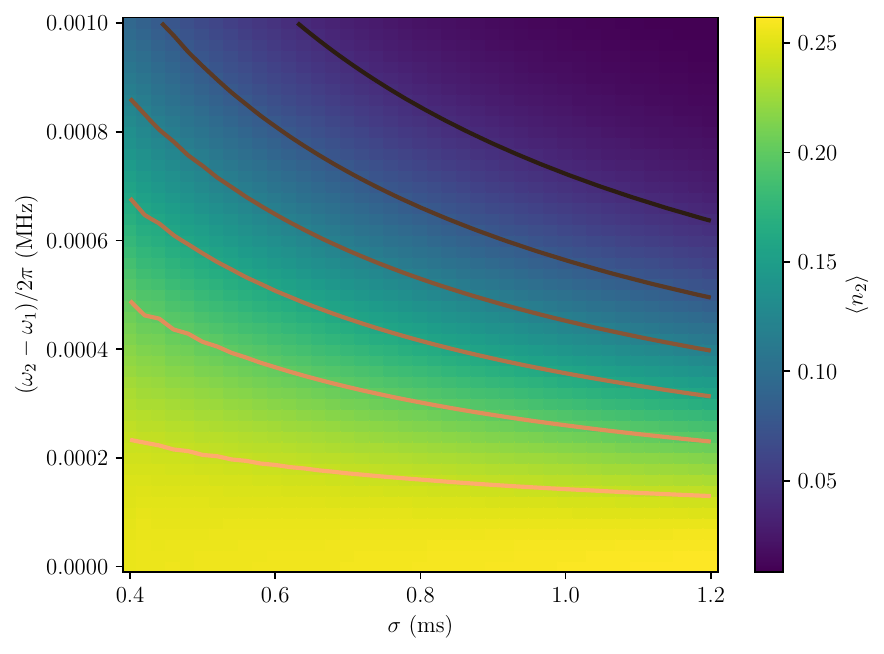}
    \caption{Higher resolution sweep close to $\delta=0$, with $\tau=\sigma/1.43$. Contours lines are included to clearly show that the behavior is approximately hyperbolic.}
    \label{sigma-omega-sweep-zoomed}
\end{figure}

\begin{figure}[h!] 
\includegraphics[width=0.99\linewidth]{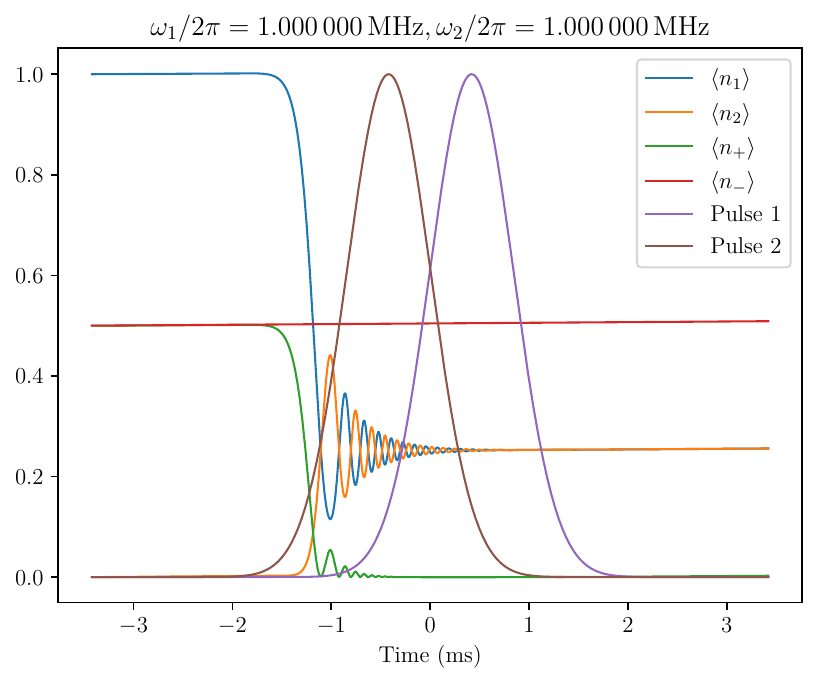}
\caption{STIRAP state transfer plotted when $\omega_1 = \omega_2$. We display $\langle n_1\rangle$, $\langle n_2\rangle$, $\langle n_+\rangle$, and $\langle n_-\rangle$, which are respectively the expectation value of the number of phonons in mode 1, mode 2 and the collective modes $+$ and $-$. We find that when the mode frequencies begin to coincide, the mechanical modes begin to behave as a collective mode, with frequency set by the amplitude of the pulses, $\alpha$. }
\label{at-diagonal}
\end{figure}

The reason for the remaining population in the mechanical modes is that,
when $\omega_1=\omega_2$, the cavity mode decouples with the collective mode
corresponding to the destroy operator $\hat b_-$
but only couples to the collective mode corresponding to the creation operator $\hat b_+$, where they can be represented as:
\begin{equation}
\label{collective without rwa}
\hat b_+=\frac{g_1\hat b_1+g_2\hat b_2}{\sqrt{g_1^2+g_2^2}},\quad
\hat b_-=\frac{g_2\hat b_1-g_1\hat b_2}{\sqrt{g_1^2+g_2^2}}.
\end{equation}
Figure~\ref{at-diagonal} shows how the expectation values of $n_\pm=b_\pm^\dagger b_\pm$ changes during the simulation when $\omega_1=\omega_2$.
We can see that $\left<n_-\right>$ does not change over time because it is not coupled to the cavity mode,
while $\left<n_+\right>$ shows the behavior of Rabi oscillation, whose amplitude decays due to cavity dissipation.
To see why only one collective mode is coupled to the cavity mode, we can rewrite the Hamiltonian in \eqref{triply} in terms of the operators associated with the collective modes:
\begin{equation}
\begin{split}
H = \sum_{i,\pm} \alpha_i & \left(\hat a^\dagger e^{i\Delta_it} + \hat ae^{-i\Delta_it}\right) \\
& \times \left(g_\pm\hat b_\pm e^{-i\omega t} + g_\pm^*\hat b_\pm^\dagger e^{i\omega t}\right)
\end{split}
\end{equation}
Here, $\omega=\left(\omega_1+\omega_2\right)/2$, $\delta=\left(\omega_1-\omega_2\right)/2$, and
\begin{equation}
g_+=\frac{g_1^2e^{-i\delta t}+g_2^2e^{i\delta t}}{\sqrt{g_1^2+g_2^2}},\quad
g_-=\frac{g_1g_2\left(e^{-i\delta t}-e^{i\delta t}\right)}{\sqrt{g_1^2+g_2^2}}.
\end{equation}
For a fixed time $t$, when $\delta$ approaches zero, $g_-$ goes to zero.

Because the simulation time interval is determined by the pulse width $\sigma$ and the pulse interval $\tau$,
the coupling $g_-$ is negligible compared to $g_+$ when $|\delta|\ll1/\tau$ or $|\delta|\ll1/\sigma$.
This condition can be confirmed by looking at Figure~\ref{sigma-omega-sweep-zoomed},
where a sweep over $\delta$ and $\sigma$ is done.
During the sweep, $\tau=\sigma/1.43$ has fixed ratio to $\sigma$
to unify both conditions $|\delta|\ll1/\tau$ and $|\delta|\ll1/\sigma$.
One can see that the equal-height contours of final values of $\left<n_2\right>$ are largely hyperbolic showing the inverse proportionality to $\sigma$.

From Figure~\ref{sweep-omega1-omega2}, we can also see peculiar behaviors near $\omega_1=3\omega_2$ and $\omega_2=3\omega_1$,
where the final values of $\left<n_2\right>$ exceeds $1$,
implying that quantum states with more than one phonon in the second mechanical mode are populated.
This can be explained using the formalism of discrete path integrals, where
the quantum time evolution operator is derived using the Picard iteration method applied to the Schrödinger equation. By mapping the Hamiltonian's matrix elements to a directed graph, it can be demonstrated that the transition amplitude between states is equivalent to the sum of all possible walks across the Hamiltonian graph.

The Schr\"odinger equation is $\frac d{dt}|\psi\rangle=-i\hat H(t)\,|\psi\rangle$,
where the Hamiltonian $H(t)$ may be time-dependent.
The full time evolution operator can be denoted as $K(t)$ so that $|\psi(t)\rangle=\hat K(t)\,|\psi(0)\rangle$ for any initial value $|\psi(0)\rangle$.

Because the Schr\"odinger equation is an ordinary differential equation, its solution can be expressed as the limit of the Picard iteration \cite{Coddington_Levinson_2012}:
\begin{equation}
    |\psi(t)\rangle=\lim_{n\to\infty}|\varphi^{(n)}(t)\rangle.
\end{equation}
The fixed-point iteration that defines the sequence $\{|\varphi^{(n)}(t)\rangle\}$ is
\begin{align}
    |\varphi^{(0)}(t)\rangle&=|\psi(0)\rangle,\\
    |\varphi^{(n+1)}(t)\rangle&=|\psi(0)\rangle-i\int_0^tdt'\,\hat H(t')\,|\varphi^{(n)}(t')\rangle.
\end{align}
By the linearity of $\hat H$, we can then convert this into a series expansion
\begin{equation}
    |\psi(t)\rangle=\sum_{n=0}^\infty|\psi^{(n)}(t)\rangle.
\end{equation}
Each term in the series is defined as
\begin{equation}
    |\psi^{(n)}(t)\rangle=|\varphi^{(n)}(t)\rangle-|\varphi^{(n-1)}(t)\rangle=\hat K^{(n)}(t)\,|\psi(0)\rangle.
\end{equation}
Here, it can be easily proven using mathematical induction that
\begin{equation}
    \hat K^{(n)}(t)=(-i)^n\int_0^tdt_n\,\hat H(t_n)\cdots\int_0^{t_2}dt_1\,\hat H(t_1).
\end{equation}
Specially, $\hat K^{(0)}(t)$ is the identity operator.
It can easily be deduced that the full time evolution operator is the sum $\hat K(t)=\sum_n\hat K^{(n)}(t)$.

Now, an orthonormal basis $\{|x\rangle\}$ is chosen,
and the matrix elements of $\hat K^{(n)}(t)$ can be expressed
in terms of the matrix elements $h_{yx}(t)=\langle y|\,\hat H(t)\,|x\rangle$ of the Hamiltonian.
\begin{widetext}
\begin{equation}
\label{K^{(n)}(t)}
    \langle y|\,\hat K^{(n)}(t)\,|x\rangle=\sum_{x_1,\ldots,x_{n-1}}
    (-i)^n\int_0^tdt_n\,h_{yx_{n-1}}(t_n)
    \int_0^{t_n}dt_{n-1}\,h_{x_{n-1}x_{n-2}}(t_{n-1})\cdots
    \int_0^{t_3}dt_2\,h_{x_2x_1}(t_2)
    \int_0^{t_2}dt_1\,h_{x_1x}(t_1).
\end{equation}
\end{widetext}
It is then beneficial to think of the basis vectors as vertices of a graph,
and the matrix element $h_{yx}(t)$ is the edge weight of a directed edge from $|x\rangle$ to $|y\rangle$.
This directed graph is sometimes referred to as the Hamiltonian graph \cite{Kalev_Hen_2025}.
Each specific choice of $x_1,\ldots,x_{n-1}$ is a choice of walk of length $n$ starting at $x$ and ending at $y$,
and the summand in Equation~\ref{K^{(n)}(t)} can be considered as a contribution from a specific choice of a walk:
\begin{equation}
\label{walk term}
\begin{split}
    &K(x_0,\ldots,x_n;t)\\&=(-i)^n\int_0^tdt_n\,h_{x_nx_{n-1}}(t_n)\cdots\int_0^{t_2}dt_1\,h_{x_1x_0}(t_1).
\end{split}
\end{equation}
it gives the full $\langle y|\,\hat K^{(n)}(t)\,|x\rangle$ when summed over all possible walks of length $n$ from $x$ to $y$.
Because of the full-time evolution $\hat K(t)=\sum_n\hat K^{(n)}(t)$,
its matrix elements are just the sum over all possible walks of all lengths:
\begin{equation}
\label{walk sum}
    \langle y|\,\hat K(t)\,|x\rangle=\sum_{\text{walks $x\to y$}}K(\text{walk};t).
\end{equation}

The matrix elements of the Hamiltonian are given by the coefficients of the eight terms
$\hat a\hat b_j$, $\hat a^\dagger\hat b_j$, $\hat a\hat b_j^\dagger$, and $\hat a^\dagger\hat b_j^\dagger$ ($j=1,2$) in the expansion of \eqref{triply}.
The four terms $\hat a^\dagger\hat b_j$ and $\hat a\hat b_j^\dagger$ are the only terms that are not quickly oscillating. By applying the rotating-wave approximation, we can eliminate other terms in the Hamiltonian, leaving us a Hamiltonian equivalent to a three-state Hamiltonian,
explicitly shown in \eqref{hamiltonian rwa}.
However, sometimes these fast oscillating terms can have resonances that make their contributions to the discrete path integral (the sum shown in \eqref{walk sum}) non-negligible,
which happens exactly when $|\Delta_2-\Delta_1|$ is equal to $2\omega_1$ or $2\omega_2$.
For example, consider the walk $w:|0,0,1\rangle\to|1,0,2\rangle\to|0,0,3\rangle$
through two hopping terms $\hat a^\dagger\hat b_2^\dagger$ and $\hat a\hat b_2^\dagger$.
According to \eqref{walk term},
the contribution of this walk to the discrete path integral is
\begin{equation}
K(w,t) = -\int_0^t dt_2 \sqrt{3} A_-(t_2)
\int_0^{t_2} dt_1 \sqrt{2} A_+(t_1),
\end{equation}
where $A_\pm(t)$ is defined for abbreviation purposes as
\begin{equation}
A_\pm(t)=g_2\left(\alpha_1e^{\pm i\Delta_1t}+\alpha_2e^{\pm i\Delta_2t}\right)e^{i\omega_2t}.
\end{equation}
If we assume that $\alpha_{1,2}$ does not depend on time
(valid because they change slowly),
the integral can be carried out analytically to retrieve a complicated but closed-form expression
with $(\Delta_1-\Delta_2-2\omega_2)(\Delta_2-\Delta_1-2\omega_2)$ in the denominator.
When $|\Delta_1-\Delta_2|=2\omega_2$, a resonance occurs such that $K(w,t)$ grows linearly in time.
A special case occurs, when $\Delta_1=3\Delta_2=3\omega_2$,
\begin{equation}
K(w,t) = \text{oscillatory}
+ i\sqrt{\frac{3}{8}} \frac{\left(\alpha_1 - 2\alpha_2\right)\alpha_2 g_2^2}{2\omega_2} t.
\end{equation}
It is then responsible for the population in the $|0,0,3\rangle$ state,
thus explaining the peculiar behavior when $\omega_1=3\omega_2$.
Similar arguments can be made for the case when $\omega_2=3\omega_1$.

\begin{figure}[!htb]
\centering
\includegraphics[width=0.90\linewidth]{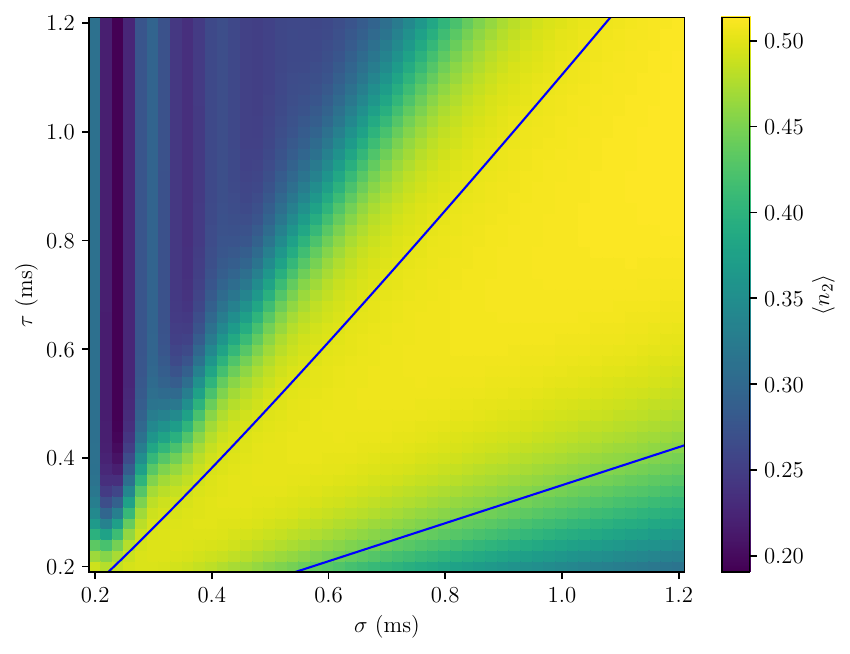}
\caption{Pulse width ($\sigma$) and pulse separation ($\tau$) versus the expectation value of the number of transferred phonons $\langle n_2\rangle$ for fSTIRAP.
The area between the blue lines is the region that satisfies the adiabaticity condition given in \eqref{adiabaticity condition}.}
\label{fig:tau sigma sweep fractional}
\end{figure}

The FSTIRAP dependence of parameters is much the same except for the conditions set by the adiabaticity condition, \eqref{adiabaticity condition}.  Figure~\ref{fig:tau sigma sweep fractional} shows the dependence of the average occupation of mechanical mode two, $\left<n_2\right>$, on $\sigma$ and $\tau$.
The blue lines displayed show the adiabaticity condition for $n=5$, which sets a region over which fSTIRAP will be adiabatic.

An important facet of the case when the environment's temperature is 1K, is the dimensionality of our system. For the results shown here, a dimensionality of DIM=5 for the mechanical states has been used. However, this means that if the steady state is considered, then  the density matrix would have a population of $\sim$1/DIM for each phonon occupation. This is not physically accurate for a system, with $\bar{n}_b\approx17000$, however $\Gamma_{mi}$ must also be considered. $\Gamma_{mi}$ modifies the thermalization present and is used to set an effective heating rate as $\Gamma_{mi} \bar{n}_b$. Provided that $G_{i,j}$ is significantly larger than this heating rate ($G_{i,j} \gg \Gamma_{mi}\bar{n}_b$), the state transfer completes well before the mechanical modes thermalize with the high-temperature bath. As $\Gamma_{mi}$ is small, the effects of $\bar{n}_b$ on our system is slow, and as a result, increasing dimensionality has little effect on the outcomes presented here. This can be checked, by examining how the fidelity in Table~\ref{tab:stirap_temp}, for the 1K cases, changes when the dimensionality is increased to $N_1=N_2=10$. A change of $\sim0.2\%$ is observed, which implies that our master equation simulation is well-converged for the parameters discussed.

\section{State Transfer: Experimental Considerations}
\label{experimental appendix}
As part of the main text's goal is to suggest an experimental protocol, additional experimental considerations must be included.

It can be assumed that a state-of-the art system can be cryogenically cooled to \SI{20}{mK} and that the membrane is heated to \SI{100}{mK} by the pump and probe driving fields \cite{Laser_Cooling,Measurement_Motional}. However, for the optomechanical parameters considered in this work, the average phonon occupation of mechanical modes at equilibrium with the environment is large. This requires us to further optically cool the mechanical modes of interest.

We begin by cooling our system, comprised of a single phononic crystal membrane and optical cavity, using a red detuned laser ($\Delta = -\omega_m$) in the resolved sideband regime ($\kappa\ll\omega_m$). This creates an interaction between the cavity and resonator mediated by \eqref{triply}. In the sideband-resolved regime, the ratio of the rates of anti-Stokes (blue-shifted) to Stokes (red-shifted) photons created by this interaction tends towards infinity, as the rate of Stokes photons created in the cavity approaches zero and the rate of anti-Stokes photons approaches the flux of thermal phonons from the environment \cite{Polzik}. This sets an optical cooling limit, for our system in the sideband-resolved regime, given by the steady-state final phonon population  \cite{Laser_Cooling,Optomechanics}:
\begin{equation}
    \bar{n}_\mathrm{f} = \frac{\Gamma_\mathrm{opt} \bar{n}_\mathrm{min}+ \bar{n}_\mathrm{th}\Gamma_\mathrm{m}}{\Gamma_\mathrm{opt}+\Gamma_\mathrm{m}}.
\end{equation}
Here $\Gamma_\mathrm m$ is the mechanical loss of the defect mode, $ \bar{n}_\mathrm{th}$ is the average thermal occupation of our environment, and $\Gamma_\mathrm{opt}$ and $\bar{n}_\mathrm{min}$ are the minimum phonon number of a mechanical mode and optomechanical damping rate, respectively, defined as \cite{Optomechanics}
\begin{equation}
\begin{split}
    \Gamma_\mathrm{opt} &= g^2\left(\frac{\kappa}{\kappa^2/4+(\Delta+\omega_\mathrm m)^2}-\frac{\kappa}{\kappa^2/4+(\Delta-\omega_\mathrm m)^2}\right)\\
    \bar{n}_\mathrm{min} &= \left(\frac{\kappa^2/4+(\Delta-\omega_\mathrm m)^2}{\kappa^2/4+(\Delta+\omega_\mathrm m)^2}-1\right)^{-1}.
\end{split}
\end{equation}
Thermal steady states of average occupation, $\bar{n}^\mathrm{r}_\mathrm{f} \approx 0.10$ \cite{Laser_Cooling,Polzik,ground_state_theory} have been shown to be experimentally achievable. For our system, steady state occupation can be demonstrated with a 5 ms red-detuned ($\Delta = -\omega_1$) constant pulse of $g^\mathrm{r} /2\pi= 1600$ Hz, which is realistic if the frequency noise of the  pump light is reduced by a narrow bandwidth filtering cavity. This corresponds to a density matrix of the form,

\begin{equation}
\label{red}
\begin{split}
    \rho_{1}^\mathrm{r} =  0.906\ket{0}\bra{0} + 0.085\ket{1}\bra{1} + \cdots \\
    \rho_{2}^\mathrm{r} =  0.906\ket{0}\bra{0} + 0.085\ket{1}\bra{1} + \cdots \\
\end{split}
\end{equation}
Where $\rho_{1}^\mathrm{r}$ corresponds to a defect mode at $\omega_1$ and $\rho_{2}^\mathrm{r}$ corresponds to a defect mode at $\omega_2$. It is clear from $\rho^\mathrm{r}$ that the single phonon occupation is much larger than the two phonon occupation.

Next, this state is converted to a close approximation of a single phonon Fock state. Detection of a heralded Stokes photon, while pumping with strong red-detuned driving fields, would project this system to the desired state \cite{Heralded_prep}, but the flux of the Stokes photons is negligible, as noted above. Pumping with a short constant blue-detuned ($\Delta = \omega_1$, $\tau_b\approx\SI{0.1}{ms}$ and $g^{b}/2\pi=\SI{2410}{Hz}$) pulse raises the occupation of our state to, $n_{\mathrm f1} \approx$ 0.20 and $n_{\mathrm f2} \approx0.12$. This corresponds to a probability of $\sim~0.1$ for a Stokes photon to be created  by mode 1 (the probability of two Stokes photons is $\sim0.1^2$). After the pulse mode 1 and 2 density matrices are given by $\rho_{1}^\mathrm b$ and $\rho_{2}^\mathrm b$,
\begin{equation}
\label{bluenum}
\begin{split}
    \rho_{1}^\mathrm{b} = 0.831\ket{0}\bra{0} + 0.140\ket{1}\bra{1} +\cdots \\
    \rho_{2}^\mathrm{b} = 0.893\ket{0}\bra{0} + 0.095\ket{1}\bra{1} +\cdots
\end{split}
\end{equation}

Once a projective measurement has been made by a single photon detector (SPD) on our state, we can collapse the vacuum amplitude, as we would deterministically know that a phonon has been created by the Stokes scattering process in mode 1. This would have a high probability of being a single phonon, as the two phonon probability amplitude is much less than the single phonon probability amplitude seen in \eqref{bluenum} \cite{Remote}. This configured density matrix would correspond to,
\begin{equation}
\label{conf}
    \rho_{1}^\mathrm{conf} = 0.831\ket{1}\bra{1} + 0.140\ket{2}\bra{2}+ \cdots
\end{equation}
The generated Stokes photon at $\omega_\mathrm{cav}$ is accompanied by strong pump fields approximately \num{e9} stronger due to low scattering probability $\sim(g_0/ \kappa)^2$. To filter out the pump light fields, the light transmitted through the cavity with the membrane is sent through a set of consecutive narrow-linewidth filtering cavities, as demonstrated in \cite{Polzik}, where \SI{150}{dB} suppression of light at \SI{1.5}{MHz} was shown.

The DCR of the detector and overall detection efficiency during heralding must be considered to provide a realistic estimate of whether experimentation is achievable. The overall detection efficiency ($\eta$) is set by the detection efficiency of the single photon detector and the optics of our system, including the filtering cavities. Similar experimentation has achieved an overall detection efficiency of 2.5\% \cite{Polzik}; however, this can be increased to 7.5\% by using superconducting nanowire SPDs. This detection efficiency of 7.5\% will be used throughout the following.

If the detector clicks due to dark counts, the state of the first mode is unchanged from $\rho_b^1$. The rate of detected Stokes photons from the heralding is \SI{75}{Hz} ($(\tau_\mathrm{b}/p\eta)^{-1}$), where $p\approx0.1$ is the probability that a single Stokes photon is created). Considering a DCR of \SI{10}{Hz} we would see an overall configured state of,

\begin{equation}
\begin{split}
      \rho_1^\mathrm{i}&= \frac{\SI{10}{Hz}}{\SI{10}{Hz}+\SI{75}{Hz}}\rho_1^\mathrm b +\frac{\SI{75}{Hz}}{\SI{10}{Hz}+\SI{75}{Hz}}\rho_1^\mathrm{conf}\\
    \rho_1^\mathrm{i}&=0.098\ket{0}\bra{0} + 0.750\ket{1}\bra{1} +\cdots  
\end{split}
\end{equation}

To read out the transferred state, we employ a short red detuned pulse ($\approx\SI{0.5}{ms}$, $g_1 / 2\pi=\SI{5000}{Hz}$) to transfer the single phonon population into anti-Stokes photons with a probability of 99.8 \% (calculated from $e^{-\Gamma_\mathrm{opt}\tau}$). This pulse duration is chosen to optimize the balance between heating effects ($\Gamma_\mathrm m\cdot n_\mathrm{th}\cdot\SI{0.5}{ms} \approx0.06$ phonons will enter into a mode by thermalization) and
the probability of converting phonon population into anti-Stokes photons. The anti-Stokes photons created through this process, would then be fiber collected and detected by superconducting SPDs (SSPD).

The DCR is a main feature of SSPDs that usually presents a challenge in low (\si{MHz}) frequency experimentation \cite{interferometry}. Conservatively, we would expect the DCR to be $\sim\SI{10}{Hz}$ for our system \cite{Superconducting_DCR,Ultimate_DCR}. On the time scale of our experiment during heralding (0.1 ms) and readout (0.5 ms), we would expect to see the DCR in combination with \SI{75}{Hz} (heralding) and \SI{15}{Hz} (readout) signals. These two signal rates are found by combining the detection efficiency with the two timescales, as shown above.

The DCR is non-negligible in comparison to the signal rates. However, in the case of the state preparation heralding the DCR is considered, and as the DCR would be independently measured and is constant, it can be subtracted from the readout signal rate. Additionally, it has been shown that by operating SSPDs with a low bias current, in combination with a cryogenically cooled optical band-pass filter, the DCR can be reduced to an almost negligible level (\num{e-4}) \cite{Ultimate_DCR}.

Timing considerations are a major challenge for low frequency experimentation \cite{interferometry}. However, the time to get a single click (using our system parameters) from blue-detuned heralding can be estimated from the time of detection, efficiency, and probability of creating a Stokes photon. 

\begin{equation}
    T_\mathrm{h}=\frac{\SI{5}{ms}+\SI{0.1}{ms}}{0.1\eta}=\SI{0.7}{s}
\end{equation}

Following a detection event of the Stokes photon, the STIRAP sequence and readout pulse are sent, which results in a final probability of detection of
\begin{equation}
    P_\mathrm f=\eta\left(1-\rho_{00}\right)<0.075.
\end{equation}
This sets a time for a click at readout of $T_\mathrm r=T_\mathrm h/P_\mathrm f >\SI{9.3}s$ depending on the density matrix to be measured.

Finally, we consider the presence of nearby mechanical modes in the membrane. STIRAP protocol relies on the heralding Stokes photons from the blue-detuned pulse and anti-Stokes photons from the read-out pulse. These two pulses affect other modes of the membrane in the frequency range of a few $\kappa$ around $\omega_1$ and $\omega_2$. As a consequence, a flux of spurious Stokes and anti-Stokes photons is produced by these modes. Modes have been observed in our membranes with quality factors as low as $10^6$ in the vicinity of mode 1 and 2. The upper bound of the rate for the phonons to enter these modes from the environment is $\sim\SI{100}{kHz}$ (based on $\Phi\approx\Gamma_\mathrm m n_\mathrm{th}$ where $n_\mathrm{th}=\num{e5}$). Additionally, the upper limit for the conversion of phonons into photons for these modes is $\Gamma_\mathrm{opt} / 2\pi<\SI{2}{kHz}$.

Because we are operating in the bandgap of our membrane, the spurious modes are at least \SI{70}{kHz} away from modes 1 and 2. This is also the minimal frequency separation of the spurious photons from $\omega_\mathrm{cav}$, where the heralded photons are detected passing through the cascade of filtering cavities described above. These filtering cavities produce at least \SI{50}{dB} of isolation at \SI{70}{kHz}, which decreases the flux of the spurious photons \num{e5} times not taking into account the detection efficiency. This causes the rate of detector clicks due to other modes in the membrane to become negligibly small and does not affect the STIRAP protocol.

\bibliographystyle{unsrt}
\bibliography{biblio}
\end{document}